\documentclass[journal]{IEEEtran}
\usepackage{cite}
\usepackage{amsmath,amssymb,amsfonts,mathrsfs}
\usepackage{algorithmic}
\usepackage{graphicx}
\usepackage{epstopdf}
\usepackage{textcomp}
\usepackage{amsmath}
\usepackage{amsthm}
\usepackage{booktabs}
\usepackage{lettrine}
\usepackage{enumitem}
\usepackage{amssymb}
\usepackage{float}
\pdfoutput=1
\usepackage{mathptmx}
\usepackage{bm}
\DeclareMathAlphabet{\mathcal}{OMS}{cmsy}{m}{n}
\begin{document}
	
	\title{Rate-Splitting Multiple Access in Multi-cell Dense Networks: A Stochastic Geometry Approach}
	\author{{Qiao Zhu, Zhihong Qian, \textit{Senior Member, IEEE}, Bruno Clerckx, \IEEEmembership{Fellow, IEEE}, and Xue Wang}	%\IEEEauthorblockA{\textit{College of Communication Engineering}, \textit{ Jilin University}, Changchun, China \\
		%	zhuqiao$\_$jlu@163.com, dr.qzh@163.com, jluwangxue@163.com}}
		% <-this % stops a space
		\thanks{This work has been funded by the National Science Foundation of China under Grant 61771219 and China Scholarship Council, and in part by the Science and Technology Research Funds of Jilin Province under Grant 20190303137SF and 20200401084GX. (\textit{Corresponding author: Xue Wang}).}
		\thanks{Qiao Zhu, Zhihong Qian, and Xue Wang are with Jilin University, Changchun, China. (e-mail: 	zhuqiao$\_$jlu@163.com, dr.qzh@163.com, txwangxue@jlu.edu.cn)}% <-this % stops a space
		\thanks{Bruno Clerckx is with Imperial College London, London SW7 2AZ, U.K. (e-mail: b.clerckx@imperial.ac.uk).}% <-this % stops a space
		
	}
	%\markboth{Journal of \LaTeX\ Class Files,~Vol.~, No.~, August~2021}%
	%{Shell \MakeLowercase{\textit{et al.}}: Bare Demo of IEEEtran.cls for IEEE Journals}

	\maketitle

	\begin{abstract}
		In this paper, the potential benefits of applying the Rate-Splitting Multiple Access (RSMA) in multi-cell dense networks are explored. Using tools of stochastic geometry, the sum-rate of RSMA-enhanced multi-cell dense networks is evaluated mathematically based on a Moment Generating Function (MGF) based framework to prove that RSMA is a general and powerful strategy for multi-antenna downlink systems. Further elaboration of the systematic performance metrics is undertaken by developing analytical expressions for area spectral efficiency and sum-rate in the RSMA-enhanced multi-cell dense networks. Based on the tractable expressions, we then offer an optimization framework for energy efficiency in terms of the number of antennas. Additionally, simulation results are shown to verify the accuracy of our analytical results and provide some insightful insights into system design. Analytically, it has been shown that: 1) the sum-rate of RSMA-enhanced multi-cell dense networks is significantly influenced by the power splitting ratio, and there is a unique value that maximizes the sum-rate; 2) the RSMA-enhanced multi-cell dense networks transmission scheme has superior sum-rate performance compared with Non-Orthogonal Multiple Access (NOMA) and Space-Division Multiple Access (SDMA) in a wide range of power splitting ratio; 3) By increasing the number of antennas and BS density in an RSMA-enhanced multi-cell dense network, the area spectral efficiency can be substantially enhanced; 4) As for energy efficiency, there exists an optimal antenna number for maximizing this performance metric. 
	\end{abstract}
	
	
	\IEEEpeerreviewmaketitle

	\section{Introduction}
	\IEEEPARstart{T}{he} proliferation of new applications and services have led to a dramatic increase in the demand for network capacity and connectivity in recent years. In the next decade (beyond 5G and 6G), wireless networks are expected to support extremely high traffic and a gigantic number of connections with minimal latency \cite{9330587}. To keep up with the exponential growth of capacity, along with the improvement of customer experiences and demands, the mobile communication technologies need to be continuously improved. This urges the development of many promising solutions such as network densification \cite{7476821}, Multiple Access (MA) \cite{9154358}, Intelligent Reflecting Surface (IRS) \cite{8741198}, massive Multiple-Input Multiple-Output (massive MIMO) \cite{8804165}, millimeter wave (mm-Wave) communication \cite{7414384} and so on. In light of these candidate technologies, network densification and multiple access are crucial to achieving massively scalable connectivity, extended coverage, and huge traffic in the next generation networks\cite{9052699}. 
	
	Dense network is characterized by the deployment of numerous Base Stations (BSs) in traffic hotspots, which generate a lot of traffic. The shift in operating mode from a network-centered mode to one that's more user-centered supports this claim. One user would generate traffic in the user-centric paradigm that will consume the whole spectrum of a BS \cite{7476821}. A higher density of small cells enhances signal quality and increases the use of the spectrum per square meter. Nevertheless, providing a system that can affordably support so many cells per unit area proves to be a cumbersome task \cite{7306534}. There is a limitation on the deployment of BSs in dense networks due to the increased network energy consumption \cite{8754736}.
	
	Combining multi-antenna and multiple access techniques is a very promising method of enhancing spatial reuse in addition to densifying the network to meet the huge throughput anticipated in the next generation networks \cite{9052699}. The well-established multiple access technology, SDMA, is responsible for numerous techniques in 4G. Orthogonal Multiple Access (OMA) superposes users at the different time-frequency resource and separates user via a proper use of the spatial dimensions. For the message to be decoded, residual interference would need to be present in all messages from other users. Channel orthogonality and strength have a significant effect on its performance. Rather than assigning users to orthogonal dimensions as in OMA, power-domain NOMA superposes users in the same time-frequency resource and discriminates them based on power characteristics \cite{7842433, 8114722}. For downlink multi-user communications, this operation is often handled using Superposition Coding (SC) on the transmit side and Successive Interference Cancellation (SIC) on the receiver side. Users who are co-scheduled are forced to fully decode the streams belonging to other users to avoid interference. Numerous efforts have been done in recent years to improve the multiplexing gain in the next generation multiple access techniques for the downlink communication system \cite{7433470, 7236924, 8352617}. While NOMA increases spectral efficiency and connectivity, it also increases the SIC operation complexity.
	
	A more general and powerful multiple access method RSMA is a form of power-domain non-orthogonal transmission strategy based on multi-antenna Rate-Splitting (RS) at the transmitter and SIC at the receiver \cite{DBLP:journals/corr/abs-1710-11018, 9451194}. The RSMA for the multi-antenna Broadcast Channel (BC) is also based on SIC and has been developed in parallel to NOMA. Through a message split at the transmitter, RSMA enables users to partially decode the message of other users so as to partially decode the interference and partially treat the remaining interference as noise. SDMA, multicasting and NOMA, on the other hand, rely exclusively on the three extremes or a combination of them and can be treated as three special cases of RSMA\cite{DBLP:journals/corr/abs-1710-11018}.
	
	The concept of RS is originally appeared in Carleial's work in 1978 \cite{1055812}, and then it is leveraged and extended in \cite{DBLP:journals/corr/abs-1710-11018, 9451194, 7152864, 7555358} for multi-antenna Broadcast Channel (BC). To partially decode messages belonging to other users, RS splits their messages into common and private parts. Multiple users encode and decode the common parts jointly, whereas only the corresponding users decode the private parts. Prior to accessing private messages, users also rely on SIC to decode common messages. Nevertheless, research has shown that NOMA with multiple SIC layers may not be as efficient and support fewer users than RSMA with only one SIC layer. A number of features of RSMA make it more robust than SDMA, multicasting and NOMA, to channel strength and direction disparities among users, inaccuracy in channel state information at the transmitter (CSIT) and network loads \cite{9451194}. 
	
	In summary, combining RSMA and network densification will provide the common ground for satisfying the various requirements of next generation access networks, where higher capacity can be achieved and a massive number of connections can be found \cite{9234747}.     
	\subsection{Motivation and State of the Art}
	Sparked by the potential benefits, we therefore explore the potential performance enhancement provided by RSMA for the dense networks. RSMA has been experiencing a new start due to the development of multiple-antenna technology and lower-complexity SIC \cite{7470942}. According to the researchers, it is expected to yield promising performance gains and complexity reduction benefits under both overloaded and underloaded scenarios \cite{7470942, DBLP:journals/corr/abs-1710-11018}. RSMA has attracted the attention of researchers recently, particularly focused on optimizing precoding and power allocation for single-cell systems. RS is generalized to massive MIMO deployments with imperfect CSIT \cite{7434643}. Particularly, a general and novel framework of Hierarchical Rate Splitting (HRS) has been proposed that is divided into an inner rate splitting and an outer rate splitting. RSMA messages are superposed with both degraded and designated beamformed streams in \cite{8019852}, which reportedly can enable maximum-minimum fairness between multiple co-channel multicast groups. Authors in \cite{8907421} have analytically demonstrated how RSMA can generalize and encompass all of SDMA, OMA, NOMA, and multicasting as special cases. Based on the results, RSMA was found to outperform NOMA and SDMA in a wide range of network loads and deployment scenarios as well as in lower complexity receiver designs than NOMA \cite{9451194}. Nevertheless, its use in multi-cell dense networks is still in its infancy and has not yet been well investigated.  
	
	Stochastic geometry is a unified mathematical paradigm for modeling the topological randomness of different types of wireless networks, characterizing their operation, and understanding their behavior \cite{haenggi2013stochastic}. As such, it can provide analytical expressions for the spatially averaged performance metrics based on the parameters and design variables of the cellular network. There have been some contributions to the field of stochastic geometry that utilize dense networks and MA techniques \cite{9052699, 7972929, 9072153}. Authors in \cite{9052699} investigate a NOMA-based technique in ultra-dense networks and derive closed-form analytical results for the uplink performance metrics. An analytical frameworks has been developed in \cite{7972929} to evaluate the probability of outages and the average achievable rate of NOMA based downlink and uplink multi-cell wireless systems. The area spectral efficiency of cluster-based MIMO-NOMA in multi-cell dense networks has also been analyzed in \cite{9072153} using tools from stochastic geometry, in which BSs are assumed to be equipped with multiple antennas to serve single-antenna users on a scheduled basis. 
	
	Although current research contributions have greatly contributed to the advancement of dense networks and MA technologies, to the best of our knowledge, no performance analysis for RSMA-enhanced multi-cell dense networks has been conducted. Performance in dense networks is mainly limited by inter-cell interference. However, most of works related to RSMA are limited to the single-cell multi-antenna BC and omit the interference caused by adjacent BSs. Limited works address RSMA in multi-cell scenarios. The authors of \cite{8732995} present a way to deal with inter-cell interference by implementing partial cooperation amongst BSs in Cloud-Radio Access Networks (C-RAN). In addition, interference-aware multi-cell cooperation has been introduced to the study of RSMA in multi-cell networks where each BS is allowed to fully cooperate with each other \cite{8756668}. It is obvious that the RSMA-assisted multi-cell dense networks is a potential architecture, which can improve the network performance significantly. However, no efforts have been devoted to systematic performance analysis on this MA technology in multi-cell dense networks. RS in the Multiple-Input Single-Output BC (MISO-BC) is well understood but is primarily focusing on communication theoretic analysis on the precoder design and max-min fairness amongst multiple co-channel multicast groups, or optimization on the energy efficiency and spectral efficiency in the single BS. Performance evaluation of large networks has not been conducted. Nevertheless, a missing piece is the development of a systematic mathematically tractable framework that can be employed to analyze and assess the spatially average performance of RSMA in multi-cell dense networks.

	\subsection{Contributions and Organization}
	
	Instead of focused on single-cell multi-antenna system, more traditional RSMA research contributions \cite{DBLP:journals/corr/abs-1710-11018, 8491100, 7434643, 8907421}, we aim to investigate RSMA-enhanced multi-cell dense networks which is considered a more challenging problem. For evaluating the spatially average performance, a mathematically tractable framework is presented to facilitate network design and optimization by tools of stochastic geometry. Downlink multi-cell dense networks are investigated in this framework, in which each BS is equipped with multiple antennas to serve multiple single-antenna users simultaneously with RSMA transmission. More specifically, the locations of BSs are modelled as a Poisson Point Process (PPP) so that inter-cell interference can be assessed. Then, the expressions that related the networks performance metrics to the system parameters and design variables are derived by stochastic geometry. Based on the proposed design, the following are the key theoretical contributions of the paper:	
	\begin{itemize}
		\item[1)]  We study a RSMA-enhanced multi-cell dense networks with perfect CSIT, and present the system modeling and performance analysis by stochastic geometry tools. This work paves the way for analyzing the spatially average performance through stochastic geometry in the use of RSMA for enhancing multi-cell dense networks. Based on this tractable model, analytical expressions for sum-rate, area spectral efficiency and energy efficiency are derived. Besides, the MGF-based approach is introduced to evaluate the performance metrics instead of its probability density function of random variables caused by topological randomness and fading channel distribution. The advantages of MGF-based approach have been demonstrated in \cite{6516171} which can not only reduce the computational complexity, but also incorporate with general fading channel distributions.
		\item[2)]  We compare RSMA with other transmission techniques. Based on our analytical and simulation results we show that the RSMA-enhanced multi-cell dense networks transmission scheme has superior sum-rate performance over both SDMA and NOMA in a wide variety of power splitting ratio, proving that it is a more efficient framework. More specifically, it is demonstrated that increasing the number of antennas will result in greater benefits than both other MA strategies.   
		\item[3)] Further, we investigate how network performance metrics are influenced by various system parameters. Specifically, the functional relationships between sum-rate and other system parameters are discussed. We found that the rate for a common stream is a decreasing function of power splitting ratio, while the sum of rates for private streams is an increasing one. Moreover, it is observed from numerical results that there exists an optimal power splitting ratio to maximize sum-rate and area spectral efficiency. 
		\item[4)] Finally, we optimize the system parameter to maximize the network energy efficiency. To evaluate the trade-off between the network energy consumption and the area spectral efficiency, a mathematical optimization problem is then formulated to optimize the energy efficiency with respect to the number of antennas.
	\end{itemize}
	
	The rest of the paper is organized as follows. In Section II, the system model for RSMA-enhanced multi-cell dense networks is introduced. In Section III, the formulas for the average common and private rates are developed, and then some of the properties of the sum-rate are analyzed analytically. In Section IV, the network capacity is shown in terms of area spectral efficiency, followed by an optimization framework. Section V presents the numerical results. This is followed by Section VI with conclusions. A summary of the symbols and notations is provided in Table I. 
	
	\begin{table}[H]	
		\caption{List of Symbols and Notations}
		\label{tab1}
		\centering
		\begin{tabular}{|l|l|} 
			\hline 
			$\Phi_b$ & PPP of BSs\\
			\hline  
			$\lambda_b$ & Density of BSs\\
			\hline 
			$\Phi_u$ & Density of users\\
			\hline
			$K$ & Number of users served by RSMA in each BS \\
			\hline  
			$N$ & Number of RSMA group in each BS \\
			\hline  
			$\left\lbrace \boldsymbol{x}\right\rbrace=\left\lbrace x_0,x_1,...\right\rbrace$ & Locations of BSs\\
			\hline  
			$M$ & Number of antennas in each BS\\
			\hline  
			$\mathcal{U}=\left\lbrace u_k\right\rbrace$ & Set of users in the same RSMA group\\
			\hline  
			$d_{0, k}, r$ & Distance from $x_0$ to $u_k$ \\
			\hline  
			$\boldsymbol{w}$ & Beamforming vectors\\
			\hline  
			$\mathrm{P}$ & BS transit power\\
			%power of common stream, power of private stream for $u_k$, \\
			\hline  
			$\alpha$ & Pathloss exponent\\
			\hline  
			$\beta$ & Power splitting ratio\\
			\hline 
			$\sigma^2$ & Noise power\\
			\hline  
			$\boldsymbol{h}$ & Fading channel\\
			\hline  
			$	\mathbb{E} \left\{ \cdot \right\} $ & Expectation operator\\
			\hline  
			$\mathcal{M} \left\{ \cdot \right\} $ & MGF of the fading channels\\
			\hline  
			$\eta$ & Signal-to-Noise-Ratio (SNR)\\
			\hline  
			${}_2F_1\left( a,b;c;d\right)$ & Hypergeometric function\\
			\hline 
			$\Gamma \left( \cdot \right) $ & Gamma function \\
			\hline  
		\end{tabular}
	\end{table}

	\section{System Model}
	
	\subsection{Network Model}
	As shown in Fig.1, we consider a single tier of multi-cell dense networks scenario that supports RSMA on downlink communication, where high network densification is introduced to enhance system performance. Then, the geographical distributions of the BSs and users are assumed to obey two independent homogeneous PPPs $\Phi_b=\left\lbrace \boldsymbol{x}\right\rbrace, \left\lbrace \boldsymbol{x}\right\rbrace=\left\lbrace x_0,x_1,...\right\rbrace $ with intensity $\lambda_b$ and $\Phi_u$ with intensity $\lambda_u$, respectively. A characteristic of this scenario is the high density of small cells with regard to the density of users. In general, each BS has a MISO-BC consisting of a transmitter with $M$ antennas and $N$ groups of users with $K$ single-antenna users served by RS in each group. Users in each group are denoted by $\mathcal{U}=\left\lbrace u_k, k=1,2,...,K\right\rbrace$. A consequence of network densification is cell size reduction, which directly translate into a decreasing number of users connected to the BS. Taking this situation into account, we assume users in different RSMA group will be assigned to orthogonal resource blocks and $K=2$ in this work \footnote{We consider two user because it captures the essence of the problem without over complicating the system model and the analysis. Extension to a larger number of users in each RSMA group is left for future work.}. Every BS serve each RSMA group with a fixed power $P$ and works in the open access mode where all users can connect to any BS. Due to the assumption that distributions of users and BSs tend to be approximately homogeneous, we can only focus on the typical BS to determine network performance metrics \cite{2018stochasticgeometry}. The concept of typical cell in stochastic geometry refers to the station $x_0=o$ located at the origin and considered under $\Phi_b$. It can be used to investigate the properties of point process and assumed to be a representative to all BSs. This assumption is widely accepted in previous works \cite{7733098}, which can also be used to identify the network-level performance of RSMA.    
	
	Without loss of generality, two fading components are considered, namely, a large-scale fading component and a small-scale fading component. For the large scale fading, we consider the standard pathloss propagation model $l\left( r \right) =\left\| r \right\| ^{-\alpha}$, where $\alpha$ is the pathloss exponent and $\left\| r \right\|$ denotes the distance between users and BSs. As the another one, all multipath fading are assumed to be independent and identically distribution (\textit{i.i.d}). 
	
	\begin{figure}[htbp]
		\centering
		\includegraphics[width=3.5in]{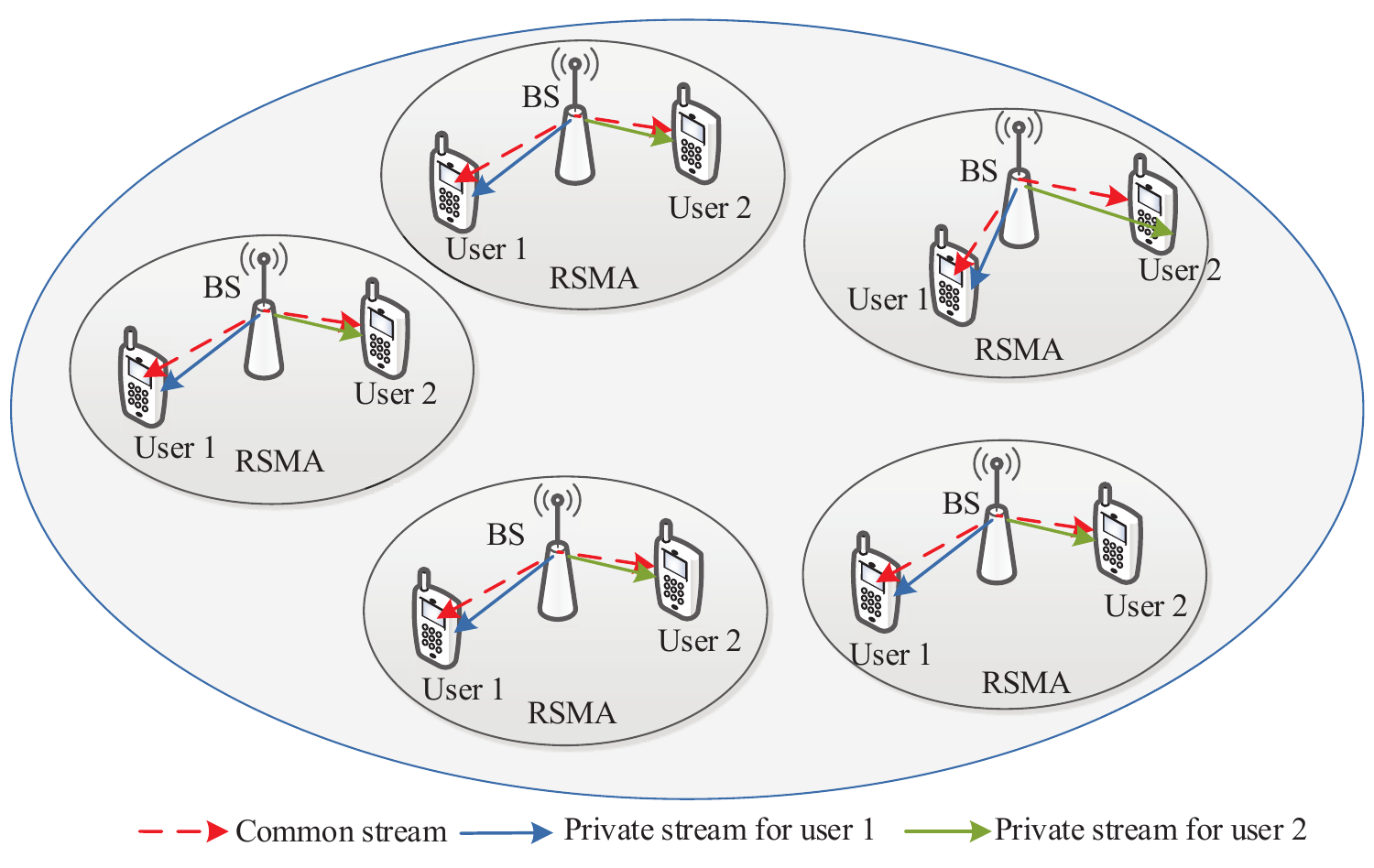}
		\caption{System model.}
		\label{fig_1}
	\end{figure}

	\subsection{RSMA}
	In each cell, the BS could serve two scheduled users at the same time based on the RSMA transmission model in this work. The RSMA strategy is pioneered by authors in \cite{9451194, 8907421}, where it is shown that this technology is a new paradigm that can meet the increasing demand for high-rate services in future networks. In this architecture, two messages $W_1$ and $W_2$ intended for $u_1$ and $u_2$ are jointly processed at the BS. More specifically, the message $W_k$ of $u_k$ could be split into a common part $W_{k,c}$ and a private part $W_{k,p}$. Then, the common parts $W_{1,c}$ and $W_{2,c}$ are combined and then encoded by a codebook shared by both users to form the common stream $s_{i, c}$ in the BS $x_i$. As a consequence, the common stream $s_{i, c}$ can be decoded by both users with zero error probability and  contains parts of the messages of $W_1$ and $W_2$. The rest parts of the messages $W_1$ and $W_2$ could be independently encoded into the private streams $s_{i, 1}$ for $u_1$ and $s_{i, 2}$ for $u_2$ in the same BS $x_i$.     The encoded signals for the two users in the BS $x_i$ are defined as $\mathbf{s}_i=\left[ s_{i,c}, s_{i,1},s_{i,2} \right] ^T$ and we follow the assumption that $\mathbb{E}\left[ \mathbf{s}_i\mathbf{s}_i^{\mathrm{H}} \right] =\mathbf{I}$. Then the private streams for both users in the same BS are superimposed over the corresponding common stream and transmitted with linear precoding. Specifically, the transmit signal of RSMA at an arbitrary BS $x_i$ can be given by 
	
	\begin{equation}
\mathbf{x}_i=\sqrt{\mathrm{P}_c}\boldsymbol{w}_{{i},{c}}s_{i,c}+\sum_{k=1}^K{\sqrt{P_k}\boldsymbol{w}_{{i},{k}}s_{i,k}}
		\tag{1}
	\end{equation}

	where $\boldsymbol{w}_{i,k}$ and $\boldsymbol{w}_{i,c}$ are the unit-norm beamforming vectors of the private stream and common stream for user $k$ in the BS $x_i$, respectively, i.e., $\left\| \boldsymbol{w}_{i,.} \right\| ^2=1$, and $\mathrm{P}_c$ is the transmit power of the common stream and $\mathrm{P}_k$ is the transmit power of the private stream $s_{k}$. For notational simplicity, we define the allocation fraction of the total transmit power P within a cluster to the private streams as $\beta\in \left( 0,1 \right]$ such that $\sum_{k=1}^K{\mathrm{P}_k}=\beta\mathrm{P}$ and then the rest of the transmit power $\left( 1-\beta \right) \mathrm{P}$ is allocated to the common stream. Moreover, we consider a uniform power allocation between the private streams such that $\mathrm{P}_1=\mathrm{P}_2=\frac{\beta\mathrm{P}}{\mathcal{K}}$. Then the transmit signal at BS $x_i$ can be rewritten as
	\begin{equation}
\mathbf{x}_i=\sqrt{\mathrm{P}}\left( \sqrt{\left( 1-\beta \right)}\boldsymbol{w}_{{i},{c}}s_{i,c}+\sum_{k=1}^K{\sqrt{\frac{\beta}{K}}\boldsymbol{w}_{{i},{k}}s_{i,k}} \right) 
		\tag{2}
	\end{equation}

	Without loss of generality, we develop the mathematical analysis based on the typical cell located at the origin \cite{haenggi2013stochastic}. It is a common concept in stochastic geometry, which refers to a typical station residing at the origin where the performance metrics of the point process can be calculated. 
	Thus, the total received signal at user $k$ served by the typical cell can be expressed as 
	\begin{equation}
		\mathbf{y}_k=d_{0,k}^{-{\alpha}/{2}}\boldsymbol{h}_{0,k}^{H}\mathbf{x}_0+\sum_{x_i\in \Phi _b\backslash  x_0}{r_{i,k}^{-{\alpha}/{2}}\boldsymbol{h}_{i,k}^{H}\mathbf{x}_i}+\boldsymbol{n}_k
		\tag{3}
	\end{equation}
	where $d_{0,k}$ is the distance between user $k$ and the serving BS $x_0$, $r_{i,k}$ is the distance between user $k$ and the interfering BS $x_i$, $\boldsymbol{n}_k\sim \mathcal{C} \mathcal{N} \left( 0,\sigma^2 \right)$ is the additive white Gaussian noise (AWGN) with equal noise variances at user $k$, 
	and $\boldsymbol{h}_{i,k}\sim \mathcal{C} \mathcal{N} \left( 0,\boldsymbol{I}_{{M}} \right)$ represents the channel vector between user $k$ and BS $x_i$, which can be written by the product of their norm and channel direction as $\boldsymbol{h}_{i,k}=\left\| \boldsymbol{h}_{i,k} \right\| \boldsymbol{\hat{h}}_{i,k}$ and assumed to be sorted in assending order, i.e., $\left\| \boldsymbol{h}_{i,1} \right\| \geqslant \left\| \boldsymbol{h}_{i,2} \right\| $. The channel matrix can be defined as $\boldsymbol{H}_i=\left[ \boldsymbol{h}_{i,1},\boldsymbol{h}_{i,2} \right]$, which are independent among BSs and of the users distances.  
	
	At each user $k$, the common stream $s_{i,c}$ is first decoded by treating the both private streams $s_{i,1}$ and $s_{i,2}$ as interference. To reconstruct and combine the signal exactly, perfect CSI at the BSs and users is considered in this paper. Thus, the SINR of the common stream is given by the 
	\begin{equation}
		SINR_{k,c}=\frac{\mathrm{P}d_{0,k}^{-{\alpha}}{\left| \boldsymbol{h}_{0,k}^{H}\boldsymbol{w}_{0,c} \right|^2}\left( 1-\beta \right)}{\sum_{j\in K}{\frac{\beta \mathrm{P}}{K}d_{0,k}^{-\alpha}\left| \boldsymbol{h}_{0,k}^{H}\boldsymbol{w}_{0,j} \right|^2} +I_c+\sigma^2}
		\tag{4} \label{SINRC}
	\end{equation}
	where $I_c=\sum_{x_i\in \Phi _b\setminus x_0}{\mathrm{P}r_{i,k}^{-{\alpha}}{\left| \boldsymbol{h}_{i,k}^{H}\boldsymbol{w}_{i,k} \right|^2}d_{0,k}^{{-\alpha}}}$ is the aggregating interference from other BSs in the multi-cell dense networks. Then based on the ideal SIC operation, the common part is subtracted from the received signal and the SINR of the two private streams can be obtained as follows
	\begin{equation}
		SINR_{k,p}=\frac{d_{0,k}^{-\alpha}\left| \boldsymbol{h}_{0,k}^{H}\boldsymbol{w}_{0,k} \right|^2\frac{\beta \mathrm{P}}{K}}{\sum_{j\in K ,j\ne k}{\frac{\beta \mathrm{P}}{K}d_{0,k}^{-\alpha}\left| \boldsymbol{h}_{0,k}^{H}\boldsymbol{w}_{0,j} \right|^2}+I_c+\sigma ^2}
		\tag{5} \label{SINRP}
	\end{equation}
	
	According to the Shannon's law, the corresponding average data rates of common part and private parts are given by  
	\begin{equation}
		\mathcal{R}_{c}=\min \left\{ \mathbb{E} \left[ \mathrm{ln}\left( 1+SINR_{1,c} \right) \right] , \mathbb{E} \left[ \mathrm{ln}\left( 1+SINR_{2,c} \right) \right] \right\} 
		\tag{6}		
	\end{equation}
	\begin{equation}
		\mathcal{R}_{k,p}=\mathbb{E}\left[ \mathrm{ln}\left( 1+SINR_{k,p} \right) \right]
		\tag{7}		
	\end{equation}
	where the minimum function is designed to ensure the common message is successfully decoded by both users and then the sum rate of the RSMA scheme can be expressed as follows
	\begin{equation}
		\mathcal{R}=\mathcal{R}_c+\sum_{k=1}^K{\mathcal{R}_{k,p}} \label{sum}
		\tag{8}	
	\end{equation}
	Assuming $C_k$ is defined as the portion of the common rate $\mathcal{R}_c$ allocated to user $k$ for transmission of common message $s_{k,c}$, then we have $\sum_{i=1}^{K}{C_k}=\mathcal{R} _c$.
	
	\subsection{Precoder Design and Distributions of Channel Gains}
	Each BS is assumed to have perfect knowledge of the CSI to its serving users, i.e., the compound matrix $\boldsymbol{H}_i\in \mathbb{C} ^{M \times K}$. To balance the system performance and implementation complexity, the private beamforming vector is obtained by zeroforcing beamforming as follows 
	\begin{equation}
		\boldsymbol{W}_i=\boldsymbol{H}_i\left( \boldsymbol{H}_i^H\boldsymbol{H}_i \right) ^{-1}=\left[ \widetilde{\boldsymbol{w}}_{\boldsymbol{i},1},\widetilde{\boldsymbol{w}}_{\boldsymbol{i},2} \right] 	
		\tag{9}	
	\end{equation}
	where $\widetilde{\boldsymbol{w}}_{\boldsymbol{i},\boldsymbol{k}}\in \mathbb{C} ^{M}$ denotes the ZF beam associate with the private stream of user $k$ in BS $i$. And then each beam is further normalized to adjust the transmission power to satisfy the constraint, namely, $\boldsymbol{w}_{\boldsymbol{i},\boldsymbol{k}}=\frac{\widetilde{\boldsymbol{w}}_{\boldsymbol{i},\boldsymbol{k}}}{\left\| \widetilde{\boldsymbol{w}}_{\boldsymbol{i},\boldsymbol{k}} \right\|}$. Obviously, the interference from private stream in the same BS is completely mitigated.   
	
	Moreover, the beamforming vector $\boldsymbol{w}_{i,c}$ is designed to maximize the achievable rate of the common stream. Based on the assumption that each BS has a knowledge of the perfect CSI and orthogonality property, the precoder $\boldsymbol{w}_{i,c}$ can be designed in the subspace $span\left( \boldsymbol{H}_i\right)$ according to the weighted matched beamforming in \cite{7434643} as follows
	\begin{equation}
		\widetilde{\boldsymbol{w}}_{\boldsymbol{i},\boldsymbol{c}}=\sum_{K}{a_k\boldsymbol{h}_{\boldsymbol{i},\boldsymbol{k}}}
		\tag{10}	
	\end{equation}
	and then the precoder $\widetilde{\boldsymbol{w}}_{\boldsymbol{i},\boldsymbol{c}}$ can be designed as 
	\begin{equation}
		\widetilde{\boldsymbol{w}}_{\boldsymbol{i},\boldsymbol{c}}=\sum_{K}{\frac{1}{\sqrt{M K}}\boldsymbol{h}_{\boldsymbol{i},\boldsymbol{k}}}
		\tag{11}	
	\end{equation}
	Further, each beam is normalized to ensure equal power assignment, i.e., $\boldsymbol{w}_{\boldsymbol{i},\boldsymbol{c}}=\frac{\widetilde{\boldsymbol{w}}_{\boldsymbol{i},\boldsymbol{c}}}{\left\| \widetilde{\boldsymbol{w}}_{\boldsymbol{i},\boldsymbol{c}} \right\|}$. 
	
	Before delving into the technical details, it is important to understand that the channel power distribution between a multi-antenna BS and a single-antenna user is affected by both transmission technique and the status of other BS \cite{6596082}.
	Therefore, it is essential to determine the distributions of equivalent channel gains of both the signal and interference. To simplify notation, we will substitute ${\left| \boldsymbol{h}_{0,k}^{H}\boldsymbol{w}_{0,k} \right|^2}$, ${\left| \boldsymbol{h}_{0,k}^{H}\boldsymbol{w}_{0,c} \right|^2}$ and ${\left| \boldsymbol{h}_{i,1}^{H}\boldsymbol{w}_{i,k} \right|^2}$ by $\boldsymbol{\psi}_{0,k}$, $\boldsymbol{\psi}_{0,c}$ and $\boldsymbol{\psi}_{i}$. From the channel model described in subsection B and Rayleigh channel fading assumption, we have $\boldsymbol{h}_{i,k}\sim \mathcal{C} \mathcal{N} \left( 0,\boldsymbol{I}_{M} \right)$. Then based on the linear precoding and the perfect CSI assumption, the equivalent channel gains of common and private streams are distributied as $\boldsymbol{\psi }_{0,k}\sim \Gamma \left( M -N +1,1 \right), \boldsymbol{\psi }_{0,c}\sim \Gamma \left( M - N +1,1 \right)$, and the interference caused by transmission from other BSs is distributed as $\boldsymbol{\psi}_{i}\sim\Gamma \left(N,1 \right)$ \cite{6596082}.

	\section{Analysis of Sum-Rate}
	In this section, our objective is to derive analytical average data rate expressions for common stream and private streams. To be more specific, distance distributions are first derived for the typical cell and user $k$, then an insightful sum-rate expression is provided using stochastic geometry concepts, as well as Rayleigh fading channels in both signals and interference links. Even though the analytical results can be more computationally tractable by mathematical platforms such as MATLAB, it is also important to verify analytical expressions by simulation, as we will demonstrate in the simulation results section.
	\subsection{Average Data Rate of Common Stream}
	It is worth noting that the point process and the fade attenuation both have significant effects on the SINR of the cluster based RSMA users. Therefore, in this subsection, the average data rates of common stream will be derived over the spatial and temporal expectation with the tools of stochastic geometry. Namely, we consider the association scheme based on the long-term averaged maximum received power. 
	
	In RSMA scheme, we assume the user $u_1$ is the near one and the user $u_2$ is the far one, i.e. $d_{0,1}<d_{0,2}$. According to the null probability of PPP, the distribution of serving distance can be expressed as $f\left( r \right) =2\pi \lambda _bre^{-\pi \lambda _br^2}$ and $r=d_{0,k}$. Then if we define two i.i.d random variables as $r_1$ and $r_2$ with PDF $f\left( r \right)$, the serving distance can be denoted as $d_{0,1}=min\left\{ r_1, r_2 \right\} $ and $d_{0,2}=max\left\{ r_1, r_2 \right\} $. Based on the Cumulative Distribution Function (CDF) of $d_{0,k}$ and differential calculation, the Probability Distribution Functions (PDF) for aforementioned serving distance of $u_1$ and $u_2$ can be in general expressed as \cite{9072153} 
	\begin{equation}
		f_1\left( r \right) =4\pi \lambda _bre^{-2\pi \lambda _br^2} \label{f1}
		\tag{12}
	\end{equation}
	\begin{equation}
		f_2\left( r \right) =4\pi \lambda _br\left( e^{-\pi \lambda _br^2}-e^{-2\pi \lambda _br^2} \right) \label{f2}
		\tag{13}
	\end{equation}
	Next, we adopt the average date rate of the common stream to evaluate the performance of the considered RS scheme. By substituting (4) into (6) and denoting $\mathcal{G} _k=\mathbb{E} \left[ \mathrm{ln}\left( 1+SINR_{k,c} \right) \right]$, we have
	\begin{equation}
		\mathcal{G} _k=\mathbb{E}\left[\mathrm{ln}\left( 1+ \frac{\mathrm{P}d_{0,k}^{-{\alpha}}{\left| \boldsymbol{h}_{0,k}^{H}\boldsymbol{w}_{0,c} \right|^2}\left( 1-\beta \right)}{\sum_{j\in K}{\frac{\beta \mathrm{P}}{K}d_{0,k}^{-\alpha}\left| \boldsymbol{h}_{0,k}^{H}\boldsymbol{w}_{0,j} \right|^2} +I_c+\sigma^2}\right) \right] \label{Gk}
		\tag{14}
	\end{equation}
	Then the general result of average data rate is evaluated under the expectation taken over both the point process $\Phi_b=\left\lbrace \boldsymbol{x}\right\rbrace$ and fade coefficients $\left\{ \boldsymbol{h}_{i,k} \right\}$ in the following Theorem. 
	
	\textbf{Theorem 1}: \textit{The average date rate of the common stream in the multi-cell dense networks with the RSMA strategy over generalized fading channels can be given by}
	\begin{equation}
		\mathcal{R} _c=\int_0^{\infty}{\frac{\left( 1-\mathcal{M} _0\left( \eta y\left( 1-\beta \right) \right) \right) \mathcal{M} _0\left(\eta y\beta \right)}{y}}\mathcal{F}_2 \left( y \right) dy \label{RC1}
		\tag{15}
	\end{equation}
	\textit{where $\eta=\frac{P}{\sigma ^2}$ is the Signal-to-Noise-Ratio (SNR) and $
		\mathcal{M} _0\left( \eta y \right) =\mathbb{E} \left\{ e^{\eta y\boldsymbol{\psi }_0} \right\}$ is the moment generating function of the fading channels from typical BS to users and}
	\begin{equation}
		\mathcal{F} _2\left( y \right) =\left( \frac{2}{\mathcal{H} _{I _c}\left( \eta y \right)}-\frac{2}{1+\mathcal{H} _{I _c}\left( \eta y \right)} \right) \left( 1-\alpha y\mathcal{D} \left( \eta y \right) \right) \label{F2}
		\tag{16}
	\end{equation}
	\begin{equation}
		\mathcal{H} _{I _c}\left( \eta y \right) =\mathcal{L} _{I _c}\left( \eta y \right) +\mathcal{M} _{I _c}\left( \eta y \right)  \label{HI}
		\tag{17}
	\end{equation}
	\begin{equation}
		\mathcal{D} \left( \eta y \right) =\int_0^{\infty}{r^{\alpha -1}\exp \left\{ -r^{\alpha}y-\pi \lambda _br^2\mathcal{H} _{I _c}\left( \eta y \right) \right\}}dr
		\tag{18}
	\end{equation}
	\begin{equation}
		\mathcal{L} _{I_c}\left( \eta y \right) =\Gamma \left( 1-\delta \right) \sum_{j=0}^{\infty}{\left( \eta y \right) ^{j+1}\mathcal{M} _{I _c}^{\left( j \right)}\left( \eta y \right) \left[ \Gamma \left( 2-\delta +j \right) \right] ^{-1}} \label{LIC}
		\tag{19}
	\end{equation}
	\begin{equation}
		\mathcal{M} _{I _c}^{\left( j \right)}\left( \eta y \right) =\mathbb{E} \left\{ \boldsymbol{\psi }_i^{j+1}e^{-\eta y\boldsymbol{\psi }_i} \right\} 
		\tag{20}
	\end{equation}
	\textit{where $\delta =\frac{2}{\alpha}$ and $\mathcal{M} _I\left( \eta y \right) =\mathbb{E} \left\{ e^{\eta y\boldsymbol{\psi }_i} \right\} $ is the moment generating function of the fading channels from interfering BSs to users.}
	\begin{proof}
		Two auxiliary variables $S_c$ and $I_p$ are introduced to denote the received signal from the common part and the interference created by the private parts, respectively. Then we have
		\begin{equation}
			\begin{split}
				S_c&=Pd_{0,k}^{{-\alpha}}{\boldsymbol{\psi}_{0,c}}\left( 1-\beta \right)\\
				I_p&=\sum_{j\in K}{\frac{\beta \mathrm{P}}{K}d_{0,k}^{-\alpha}\boldsymbol{\psi }_{0,k}}	
			\end{split}
			\tag{21}
		\end{equation}
		Therefore, the expression for $\mathcal{G} _k$ can be rewritten as:
		\begin{equation}	
			\mathcal{G} _k=\mathbb{E}_r\left[ \mathbb{E}\left[ \mathrm{ln}\left( 1+\frac{X}{Y +Z+1}\right) \Bigg \vert r\right]  \right] 
			\tag{22}
		\end{equation}
		where $X=\frac{S_c}{\sigma^2}$, $Y=\frac{I_p}{\sigma^2}$ and $Z=\frac{I_c}{\sigma^2}$ are arbitrary non-negative random variables. Then by using the useful equation from \cite{5407601}, 
		\begin{equation}
			\mathrm{ln}\left( 1+x \right) =\int_0^{\infty}{\frac{1}{s}\left( 1-e^{-sx} \right)}e^{-s}ds
			\tag{23}
		\end{equation}
		where $x>0$, we obtain
		\begin{equation}
			\begin{split}
				\mathcal{G} _k\left(r\right)&=\mathbb{E}\left[ \mathrm{ln}\left( 1+\frac{X}{Y +Z+1}\right) \Bigg \vert r\right]\\
				&= \int_0^{\infty}\frac{e^{-z}}{z}\left( \mathcal{M} _{Y,Z}-\mathcal{M} _{X,Y,Z} \right)dz 
				\\
				&\overset{\left( a \right)}{=} \int_0^{\infty}\frac{e^{-z}}{z}\left( 1-\mathcal{M} _X \right) \mathcal{M} _Y\mathcal{M} _Zdz
			\end{split}
			\tag{24}
		\end{equation}
		where (a) follows from the random variables $X,Y,Z$ are independent and $
		\mathcal{M} _{X,Y,Z}=\mathbb{E} \left\{ e^{z\left( X +Y +Z \right)} \right\}$ is the moment generating function of random variable $X +Y +Z$. And then the expectation in (\ref{Gk}) can be derived as follows 
		\begin{equation}
			\begin{split}
				\mathcal{G} _k\left(r\right)=\int_0^{\infty}{\frac{e^{-z}}{z}}&\left( 1-\mathcal{M} _0\left( \eta r^{-\alpha}z\left( 1-\beta \right) \right) \right)\\
				& \hspace*{1cm} \mathcal{M} _0\left( \eta r^{-\alpha}z\beta  \right) \mathcal{M} _{I_c}\left( \eta z,r \right) dz \label{GKr}
			\end{split}
			\tag{25}
		\end{equation}
		where $\eta =\frac{P}{\sigma ^2}$ is the SNR, and $\mathcal{M} _0\left( \cdot \right) $ is the moment generating function of the equivalent fading channel. $\mathcal{M} _{{I}_c}\left( \eta z,r \right)$ is the moment generating function of the aggregate interference from all BSs except the serving one. According to the result in \cite{6516171}, $\mathcal{M} _{{I}_c}\left( \eta z,r \right)$ can be given as follows
		\begin{equation}
			\begin{split}
				\mathcal{M} _{{I}_c}\left( \eta z,r \right) &=\exp \left\{ \pi \lambda _br^2 \right\} \exp \left\{ -\pi \lambda _br^2\mathcal{M} _{{I}_c}\left( \eta zr^{-\alpha} \right) \right\} \\
				&\times \exp \left\{ -\pi \lambda _br^2\mathcal{L} _{{I}_c}\left( \eta zr^{-\alpha} \right) \right\}\\ 
				&=\exp \left\{ \pi \lambda _br^2\left( 1-\mathcal{H} _{{I} _c}\left( \eta zr^{-\alpha} \right) \right) \right\}  \label{MIC}
			\end{split}
			\tag{26}
		\end{equation}
		where $\mathcal{H} _{{I} _c}\left( \eta zr^{-\alpha} \right) =\mathcal{M} _{{I} _c}\left( \eta zr^{-\alpha} \right) +\mathcal{L} _{{I} _c}\left( \eta zr^{-\alpha} \right) $.
		Now we can compute the expression of $\mathcal{G} _k$ as follows
		\begin{equation}	
			\mathcal{G} _k=\int_0^{\infty}{\mathcal{G} _k\left( r \right)}f_k\left( r \right) dr \label{GK}
			\tag{27}
		\end{equation}
		Substituting (\ref{MIC}) in (\ref{GKr}) and then substituting (\ref{GKr}) along with (\ref{f1}) and (\ref{f2}) in (\ref{GK}) respectively. For simplicity of notation, a change of variables $y=zr^{-\alpha}$ is used, and an integration by parts method is applied to calculate the integral (\ref{GK}) in $r$. Then we have
		\begin{equation}
			\begin{split}
				\mathcal{R} _{1, c}=\int_0^{\infty}{\frac{\left( 1-\mathcal{M} _0\left( \eta y\left( 1-\beta \right) \right) \right) \mathcal{M} _0\left(\eta y\beta \right)}{y}}\mathcal{F}_1 \left( y \right) dy\\
				\mathcal{R} _{2, c}=\int_0^{\infty}{\frac{\left( 1-\mathcal{M} _0\left( \eta y\left( 1-\beta \right) \right) \right) \mathcal{M} _0\left(\eta y\beta \right)}{y}}\mathcal{F}_2 \left( y \right) dy
			\end{split}
			\tag{28}
		\end{equation}	
		where
		\begin{equation} 
			\mathcal{F} _1\left( y \right) =\frac{2}{1+\mathcal{H} _{{I} _c}\left( \eta y \right)}\left( 1-\alpha y\mathcal{D} \left( \eta y \right) \right)  \label{F1}
			\tag{29}
		\end{equation}
		At last, the minimum function in (6) is used to obtain the result with $\mathcal{F} _1\left( y \right) > \mathcal{F} _2\left( y \right) $. This concludes the proof.  
	\end{proof}
	
	\subsection{Average Data Rates of Private Streams}
	In this subsection, the average date rates of private parts are computed by averaging the rates over the fading channels and spatial locations of typical BS and interfering BSs. Assuming that common part can be decoded without error and the interference between private parts can be cancelled completely, the average date rates of private part for $u_1$ and $u_2$ can be expressed by  
	\begin{equation}
		\mathcal{R}_{k,p}=\mathbb{E}\left[\mathrm{ln}\left( 1+ \frac{\frac{\beta\mathrm{P} }{{K}}d_{0,k}^{-{\alpha}}{\left| \boldsymbol{h}_{0,k}^{H}\boldsymbol{w}_{0,k} \right|^2}}{{I}_k+{I}_c+\sigma^2}\right) \right] 
		\tag{30}
	\end{equation}
	where ${I}_k=\sum_{j\in K ,j\ne k}{\frac{\beta \mathrm{P}}{K}d_{0,k}^{-\alpha}\left| \boldsymbol{h}_{0,k}^{H}\boldsymbol{w}_{0,j} \right|^2}$.
	Following the similar steps in the evaluation of average rates of the common part, expressions for average date rates of private parts can be given by the following theorem. These formulations are given in terms of power allocation coefficient and fading channels.
	
	\textbf{Theorem 2}: \textit{The average date rate of the private streams in the multi-cell dense networks with the RSMA strategy over generalized fading channels can be given by}
	\begin{equation}
		\begin{split}
			\mathcal{R} _{1,p}&=\int_0^{\infty}{\frac{\left( 1-\mathcal{M} _0\left( \frac{\eta y\beta }{{K}}\right) \right) \mathcal{M} _0\left( \frac{\eta y\beta }{{K}}\right)}{y}}\mathcal{F} _1\left( y \right) dy\\
			\mathcal{R} _{2,p}&=\int_0^{\infty}{\frac{\left( 1-\mathcal{M} _0\left( \frac{\eta y\beta }{{K}} \right) \right) \mathcal{M} _0\left( \frac{\eta y\beta }{{K}}\right)}{y}}\mathcal{F} _2\left( y \right) dy \label{RP1}
		\end{split}
		\tag{31}
	\end{equation}
	\textit{where $\mathcal{F} _1\left( y \right) $and $\mathcal{F} _2\left( y \right) $ is given by (\ref{F1}) and (\ref{F2}).}
	
	\begin{proof}
		It is the same steps we take in proving Theorem 2 that are used in proving Theorem 1.
	\end{proof}

	Then by substituting the expressions of $\mathcal{R} _{c}$ and $\mathcal{R} _{k,p}$ into (8), we obtain the desired sum-rate formulation. While the expression is not closed, it can still be numerically evaluated efficiently, unlike the typical Monte Carlo methods that employ repeated sampling to estimate the network performance. Additionally, closed-form expressions of $\mathcal{M} _0\left( \cdot \right) $ and $\mathcal{M} _I\left( \cdot \right) $ for many fading channel models are provided in \cite{5238289}, \cite{6226905}, and \cite{6152071}. Then the sum-rate can be directly calculated from the MGF of signal and interference links. From the formulation of sum-rate, we can draw some important conclusions about how the system performance depends on the power allocation coefficient $\beta$, as summarized in Corollary 1.
	
	\textbf{Corollary 1}: \textit{ For Rayleigh fading, it is obviously that $\mathcal{R} _{c}$ in (15) decreases with the power splitting ratio $\beta$, while  $\sum_{k=1}^{K}{\mathcal{R}_{k,p}}$ is a strictly monotonically increasing function of $\beta$. Thus, the optimal power allocation coefficient can be found by the first-order derivative condition to maximize the sum-rate in (8). }
	
	\begin{proof}
		For $\mathcal{M} _0\left( a \right) =\frac{1}{\left( 1+a \right) ^{\zeta}}$ with $\zeta={M} -{N} +1$,  differentiating (\ref{RC1}) with respect to $\beta$, we have
		\begin{equation} 
			\frac{\partial \mathcal{R} _c}{\partial \beta}=\int_0^{\infty}{f_c\left( \beta , y \right)}\frac{\mathcal{F} _2\left( y \right)}{y}dy
			\tag{32}
		\end{equation}
		where $f_c\left( \beta , y \right)$ is given by
		\begin{equation}
			\begin{split}
				f_c\left( \beta , y \right) &=	\frac{\xi \eta y}{K}\left( 1+\frac{\eta y\beta}{K} \right) ^{\xi -1}\left( \left( 1+\left( 1-\beta \right) \eta y \right) ^{-\xi}-1 \right)
				\\
				& -\xi \eta y\left( 1+\frac{\eta y\beta}{K} \right) ^{\xi}\left( 1+\left( 1-\beta \right) \eta y \right) ^{-\xi -1}
			\end{split}
			\tag{33}
		\end{equation}	
		where $0<\left( 1+\left( 1-\beta \right) \eta y \right) ^{-\xi}\leqslant 1$ and $0<\left( 1+\frac{\eta y\beta}{K} \right) ^{\xi}\leqslant 1$. Thus we have $\frac{\partial \mathcal{R} _c}{\partial \beta}<0$. As a result, $\mathcal{R} _{c}$ is a decreasing function when $\beta \in \left( 0,1 \right) $. Similarly, Taking the first-order derivative of the rates of private streams, we have
		\begin{equation} 
			\frac{\partial \left( \sum_{k=1}^{{K}}{\mathcal{R} _{k,p}} \right)}{\partial \beta}=\int_0^{\infty}{f_p\left( \beta , y \right)}\frac{\mathcal{F} _1\left( y \right) +\mathcal{F} _2\left( y \right)}{y}dy
			\tag{34}
		\end{equation}	
		where $f_p\left( \beta , y \right)$ can be written as
		\begin{equation}
			\begin{split}
				f_p\left( \beta , y \right) =\frac{\xi \eta y}{K}\left( 1+\frac{\eta y\beta}{K} \right) ^{-2\xi -1}\left( 2-\left( 1+\frac{\eta y\beta}{K} \right) ^{\xi} \right) 
			\end{split}
			\tag{35}
		\end{equation}
		With $0<\left( 1+\frac{\eta y\beta}{K} \right) ^{-2\xi -1}<1$, it is easy to verify that $\frac{\partial \left( \sum_{k=1}^{{K}}{\mathcal{R} _{k,p}} \right)}{\partial \beta}>0$ when $\beta \in \left( 0,1 \right) $. Thus, $\sum_{k=1}^{{K}}{\mathcal{R} _{k,p}}$ is an increasing function of $\beta$. In section V, we will also look at the numerical simulation results to confirm that conclusion.
	\end{proof}

	\subsection{Sum-Rate}
	A general distribution of channel fading is considered in Theorem 1 and Theorem 2.  In consequence, the sum-rate of RSMA-enhanced multi-cell dense networks for general signal and interference fading channels can be calculated. The fading channel can thus be used for the signal link as it is suitable for dense network environments in which the serving BS are close to the active users. The solution is simplified in dense propagation networks where Rayleigh fading channels are assumed. In the following theorem, we consider a Rayleigh fading distribution of the useful link, which results in a more tractable solution.
	
	\textbf{Theorem 3}: \textit{Let the interference links and signal links experience Rayleigh fading in the propagation channels. Accordingly, the equivalent channel gains can be given by section $\uppercase\expandafter{\romannumeral2}.C$. Then, $\mathcal{L} _{{I} _c}\left( \eta y \right)$ in (\ref{LIC}) has closed-form expression as follows}
	\begin{equation} 
		\mathcal{L} _{{I} _c}\left( \eta y \right) =\frac{\eta y}{\left( 1-\delta \right) \left( 1+\eta y \right)}{}_{2}F_1\left( 2, 1, 2-\delta , \frac{\eta y}{\eta y+1} \right) 
		\tag{36}
	\end{equation}
	\textit{where ${}_2F_1\left( a,b;c;d\right)$ is the Hypergeometric function.}
	\textit{Then, the sum-rate of RSMA in multi-cell dense networks can be given by }
	\begin{equation}
		\begin{split}
			&\mathcal{R} =\int_0^{\infty}{\left( 1-\mathcal{M} _0\left( \eta y\left( 1-\beta \right) \right) \right) \mathcal{M} _0\left( {\eta y\beta} \right) \mathcal{F} _2\left( y \right)}+\\
			&\left( 1-\mathcal{M} _0\left( \frac{\eta y\beta}{{K}} \right) \right) \mathcal{M} _0\left( \frac{\eta y\beta}{{K}} \right)\left( \mathcal{F} _1\left( y \right) +\mathcal{F} _2\left( y \right) \right) \frac{dy}{y}
		\end{split} \label{RC}
		\tag{37}
	\end{equation}

	\begin{proof}
		Based on the result in \cite{6516171}, the closed-form expression of $\mathcal{L} _{{I} _c}\left( \eta y \right)$ is available by setting the Gamma distribution parameters as (1,1), which we denote in section  $\uppercase\expandafter{\romannumeral2}.C$. And then the result can be given by some algebraic manipulations. This concludes the proof.
	\end{proof}
	It is evident from the formulation of the sum-rate in (36) that the function $\mathcal{F} _k\left( y \right)$ is essential to the characterization of the network performance. Due to the fact that the integrated function of $\mathcal{F} _k\left( y \right)$ is always greater than zero, $\mathcal{R}$ is a monotonically increasing function of $\lambda _b$.
	
	For the special case of $\alpha=4$, the expression of $\mathcal{F} _k\left( y \right)$ can be further simplified to closed-form in Corollary 2.
	
	\textbf{Corollary 2}: \textit{As $\alpha=4$, the formulation of $\mathcal{F} _k\left( y \right)$ can be evaluated in closed-form as follows
		\begin{equation}
			\mathcal{F} _1\left( y \right)_{\alpha=4} =\frac{\pi \lambda _b\mathcal{H} _{{I} _c}\left( \eta y \right)}{4y\left( 1+\mathcal{H} _{{I} _c}\left( \eta y \right) \right)}\mathcal{O} \left( y \right) \label{F11}
			\tag{38}
		\end{equation}
		\begin{equation}
			\mathcal{F} _2\left( y \right)_{\alpha=4} =\frac{\pi \lambda _b}{4y\left( 1+\mathcal{H} _{{I} _c}\left( \eta y \right) \right)}\mathcal{O} \left( y \right) \label{F21}
			\tag{39}
		\end{equation}
		where
		\begin{equation}
			\mathcal{O} \left( y \right) =\sqrt{\frac{\pi}{y}}\exp \left\{ \frac{\left( \pi \lambda _b\mathcal{H} _{{I} _c}\left( \eta y \right) \right) ^2}{4y} \right\} erfc\left( \frac{\pi \lambda _b\mathcal{H} _{{I} _c}\left( \eta y \right)}{2\sqrt{y}} \right) 
			\tag{40}
		\end{equation}}
		
		\begin{proof}
			Equation (\ref{F11}) and (\ref{F21}) follows from the notable integral in \cite{Integrals} and some algebraic manipulations. This concludes the proof.
		\end{proof}
		As a consequence, given the numerical calculation of $\mathcal{M} _0\left( \cdot \right)$ with reference to the channel fading distributions, the sum-rate can be evaluated easily by software programs. 
		
		Thermal noise is not a major concern in most dense networks. It can be ignored inside the cell since it is very small compared to the desired signal power (high SNR), i.e., the interference-limited scenario. In the case that $\sigma ^2\rightarrow 0$, the following corollary specializes sum-rate to interference-limited dense networks.
		
		\textbf{Corollary 3}: \textit{As $\sigma ^2=0$, the formulation of sum-rate in (\ref{sum}) along with (\ref{RC1}) and (\ref{RP1}) can be simplified as follows}
		\begin{equation}
			\begin{split}
				\mathcal{R} =\mathcal{R} _{\eta \rightarrow \infty}=&\int_0^{\infty}{\frac{2\left( 1-\mathcal{M} _0\left( t\left( 1-\beta \right) \right) \right) \mathcal{M} _0\left({t\beta} \right)}{\mathcal{H} _{{I} _c}\left( t \right) \left( 1+\mathcal{H} _{{I} _c}\left( t \right) \right) }}\\
				&\hspace*{0.5cm} {+\frac{2 \left( 1-\mathcal{M} _0\left( \frac{t\beta}{{K}} \right) \right)\mathcal{M} _0\left( \frac{t\beta}{{K}} \right)}{\mathcal{H} _{{I} _c}\left( t \right)}}\frac{dt}{t}
			\end{split}
			\tag{41}
		\end{equation}
		
		\begin{proof}
			Manipulating $\mathcal{F} _k\left( y \right)$ by using the change of variables $t=\eta y$, and taking the limit as $\eta \rightarrow \infty$, it follows that $\mathcal{F} _1\left( \eta ^{-1}t \right) =\frac{2}{1+\mathcal{H} _{{I} _c}\left( t \right)}$ and $\mathcal{F} _2\left( \eta ^{-1}t \right) =\frac{2}{\mathcal{H} _{{I} _c}\left( t \right) \left( 1+\mathcal{H} _{{I} _c}\left( t \right) \right)}$, which completes the proof. 
		\end{proof}
		It is interesting to highlight that the sum-rate in the interference-limited networks is independent of the deployment density of BSs and transmit power. Therefore, a higher BSs density or a higher transmit power will not be effective in the improvement of sum-rate regardless of the fading channels. The conclusions in this study are in agreement with previous observations in interference-limited networks, first published in \cite{6042301}. 
		
		\subsection{Comparison with SDMA and NOMA}
		As an additional analysis to help provide a more in-depth understanding about the benefits of RS as an access paradigm that includes SDMA and NOMA as special cases, we also evaluate the average downlink rates for SDMA and NOMA in Corollary 4 and Corollary 5. Similarly, the uniform power allocation between $P_1$ and $P_2$ has been adapted to maintain consistency.
		
		\textbf{Corollary 4}: \textit{For SDMA, both private messages are encoded directly into private streams and thus no power will be allocated to the common streams ($P_c=0$) such that $\beta =1$. Then the sum-rate of $u_1$ and $u_2$ for SDMA is given by}
		\begin{equation}
			\mathcal{R} ^{SDMA}=\int_0^{\infty}{\left( 1-\mathcal{M} _0\left( \frac{\eta y}{K} \right) \right) \mathcal{M} _0\left( \frac{\eta y}{K} \right) \left( \mathcal{F} _1\left( y \right) +\mathcal{F} _2\left( y \right) \right)}\frac{dy}{y}
			\tag{42}
		\end{equation}
		
		\begin{proof}
			The proof follows by using the same steps as in sum-rate of RSMA and by taking into account that $\beta =1$.
		\end{proof}
		
		\textbf{Corollary 5}: \textit{For NOMA, the private part of $u_2$ is entirely encoded into the common stream, and then all power for the private streams is allocated to $u_1$, namely, $P_2=0$. Then the sum-rate of $u_1$ and $u_2$ for NOMA case is expressed as }
		\begin{equation}
			\begin{split}
				\mathcal{R} ^{NOMA}&=\int_0^{\infty}{\left( 1-\mathcal{M} _0\left( \eta y\left( 1-\beta \right) \right) \right) \mathcal{M} _0\left( \eta y\beta \right) \mathcal{F} _2\left( y \right)} \\
				&+\left( 1-\mathcal{M} _0\left( \eta y\beta \right) \right) \mathcal{M} _0\left( \eta y\left( 1-\beta \right) \right) \mathcal{F} _1\left( y \right) \frac{dy}{y}
			\end{split}
			\tag{43}
		\end{equation}

		\begin{proof}
			The proof follows directly from (\ref{RC1}) and (\ref{RP1}) with the assumption that all power for private streams is allocated to $u_1$. And the message of $u_2$ is encoded as common stream. That means $u_1$ must decode the message of $u_2$ and subtracted it, and then obtain its own original message. 
		\end{proof}
		
		As the BS density in multi-cell dense networks is typically large, we can ignore the noise and focus on the interference-limited regime. Next, the following two Propositions show the difference between the sum-rates of RSMA and other two multiple access schemes described in Corollary 4 and Corollary 5.
		
		\textbf{Proposition 1}: \textit{As $\sigma ^2=0$, the difference between sum-rates of RSMA and SDMA can be calculated by}		
		\begin{equation}
			\begin{split}
				&\varDelta \mathcal{R} ^{RS-SDMA}=\int_0^{\infty}{\frac{2\left( 1-\mathcal{M} _0\left( t\left( 1-\beta \right) \right) \right) \mathcal{M} _0\left( t\beta \right)}{\mathcal{H} _{I_c}\left( t \right) \left( 1+\mathcal{H} _{I_c}\left( t \right) \right)}}+
				\\
				&\frac{2\left\{ \left( 1-\mathcal{M} _0\left( \frac{t\beta}{K} \right) \right) \mathcal{M} _0\left( \frac{t\beta}{K} \right) -\left( 1-\mathcal{M} _0\left( \frac{t}{K} \right) \right) \mathcal{M} _0\left( \frac{t}{K} \right) \right\}}{\mathcal{H} _{I_c}\left( t \right)}\frac{dt}{t}
			\end{split}
			\tag{44}
		\end{equation}

		\textbf{Proposition 2}: \textit{As $\sigma ^2=0$, by comparing the sum-rates of RSMA and NOMA, the difference is given by}
		\begin{equation}
			\begin{split}
				\varDelta \mathcal{R} ^{RS-NOMA}&=\int_0^{\infty}{\frac{2\left( 1-\mathcal{M} _0\left( \frac{t\beta}{K} \right) \right) \mathcal{M} _0\left( \frac{t\beta}{K} \right)}{\mathcal{H} _{I_c}\left( t \right)}} \\
				&-\frac{2\left( 1-\mathcal{M} _0\left( t\beta \right) \right) \mathcal{M} _0\left( t\left( 1-\beta \right) \right)}{1+\mathcal{H} _{I_c}\left( t \right)}\frac{dt}{t}
			\end{split}
			\tag{45}
		\end{equation}

		\begin{proof}
			Based on the result in Corollary 3, Corollary 4 and Corollary 5, Proposition 1 and Proposition 2 can be obtained by some algebraic manipulations.  
		\end{proof}
		
		By evaluating the numerical simulation results in section V according to these propositions, we will also show that RSMA performs better than the two other MA techniques when analyzing a wide range of power splitting ratio.

		\section{Area Spectral Efficiency and Energy Efficiency}
		In this section, we first investigate the behavior of the network-level performance metric area spectral efficiency regarding different system parameters as described in the RSMA-enhanced multi-cell dense networks. Following that, we will formulate the energy efficiency maximization problem and provide the optimization framework to solve it.
		\subsection{Area Spectral Efficiency}
		In computing the network capacity, area spectral efficiency is an incredibly common performance metric that is defined by the number of transmitted nats per second per Hz per unit area. As a result of the interpretation from previous works \cite{7412737, 7438738} and the combination of our system configuration, in this work, area spectral efficiency is defined as the product of BS density and the average downlink sum-rate. Using the sum-rate analytical results in the previous section, the expression for area spectral efficiency can be given by
		\begin{equation}
			\mathcal{A} _{\mathcal{S} \mathcal{E}}\triangleq \lambda _b\mathcal{R} \label{ASE}
			\tag{46}
		\end{equation} 
		where we assume that no BS is in the idle state, i.e., all BSs in the multi-cell dense networks are actively serving a cluster of users. Obviously, area spectral efficiency is an increasing function of the BS density and the average sum-rate. Therefore, although densifying BSs does not contribute to increasing the average sum-rate in the RSMA-enhanced multi-cell dense networks, it is an effective means of meeting the explosive growth in network capacity \cite{7931666}. To better understand how area spectral efficiency responds to different $\mathcal{R}$ related to various system parameters, we will primarily investigate the impact of power allocation coefficients $\beta$ and the number of BS antennas ${M}$.
		
		Under the assumption that the BS density $\lambda _b$ is constant and the power splitting ratio $\beta$ is fixed, the relationship between the efficiency parameters area spectral efficiency and the number of BS antennas is evaluated using mathematical functions of ${M}$. Then, we have the following result to show how area spectral efficiency is impacted by variation in the number of antennas.
		
		\textbf{Remark 1}: \textit{With fixed BS density and power splitting ratio, the only parameter that the designer can adjust to improve area spectral efficiency is to increase the number of antennas ${M}$. Further, area spectral efficiency $\mathcal{A} _{\mathcal{S} \mathcal{E}}$ is an increasing and concave function of ${M}$}
		
		\begin{proof}
			If $\lambda _b$ and $\beta$ are fixed, what we need to do is to prove that $\mathcal{R}$ is an increasing function of ${M}$. It is obviously that $\mathcal{M} _0\left( a \right) =\frac{1}{\left( 1+a \right) ^{\zeta}}$ is an decreasing and concave function of ${M}$. Meanwhile, it is not difficult to prove that $\mathcal{R}$ is also a decreasing function of the MGF. Since it is increasing if the composite function has the same monotony, we conclude that $\mathcal{R}$ is an increasing function of ${M}$. The deployment of more antennas will substantially improve network capacity while the gains are diminishing, as described in \cite{7442903}.
		\end{proof}
		As shown in the definition of area spectral efficiency, $\mathcal{A} _{\mathcal{S} \mathcal{E}}$ increases linearly with the sum-rate. If the BS density and antenna count are fixed, the next step is to study how the power splitting ratio affects the network capacity through sum-rate. According to the discussion in Corollary 1, it has shown that the power allocation coefficient $\beta$ has different effects on the common rate and the sum of private rates. To analyze the network capacity, it is of vital importance to investigate how the sum-rate are impacted by variations in power allocation coefficient $\beta$. To illustrate how area spectral efficiency is affected by changes in power splitting ratio, we have the following result.

		\textbf{Remark 2}: \textit{If the BS density and number of antennas are fixed, the only parameter that can be adjusted to improve ASE is the power splitting ratio $\beta$. Further, area spectral efficiency $\mathcal{A} _{\mathcal{S} \mathcal{E}}$ is an increasing function of $\beta$}.
		
		Since this proof is like that for Remark 1, we will omit it here and have it validated in the numerical simulation results section.

		\subsection{Energy Efficiency}
		We have investigated the spectral performance of the RSMA-enhanced multi-cell dense networks in the previous subsection. Despite the benefits that come from increasing the number of antennas, the cost associated with these gains is also important to think about from the perspective of the network designer. Thus, we formulate the problem of energy efficiency maximization in terms of antenna number ${M}$. According to \cite{7442903}, network energy consumption is defined as the average energy consumption per unit area and summarized as follows:
		\begin{equation}
			E=\lambda _b\left( \frac{P}{\theta}+{M} P_{cir}+{K} ^3P_{pre}+P_0 \right) 
			\tag{47}
		\end{equation}
		where $\theta$ is the power amplifier efficiency, $P_{cir}$ denotes the circuit power per antenna, ${K} ^3P_{pre}$ represents the energy consumption for precoding, which is related to the number of users, and $P_0$ indicates the non-transmission power. It can be observed that the area spectral efficiency and energy consumption model are monotonically increasing functions of the antenna number. From an energy consumption perspective, the cost-benefit analysis can be formulated in terms of energy efficiency. It is determined by the number of bits that can be transmitted per unit use of available spectrum at the expense of one Joule in one second. Therefore, based on the definition of area spectral efficiency and the network energy consumption model, the energy efficiency maximization problem of the RSMA-enhanced multi-cell dense networks can be expressed as
		\begin{equation} 
			\begin{matrix}
				\mathbf{P}1:&		\underset{{M} \in \mathbf{N}^+}{\max}&		\frac{\mathcal{R}}{\frac{P}{\theta}+{M} P_{cir}+{K} ^3P_{pre}+P_0}\\ 
				&		s.t.&		{M} \geqslant {K}\\
			\end{matrix}
			\tag{48} \label{EE}
		\end{equation}
		where $\mathcal{R}$ is given in (\ref{RC}). In order to solve the optimization problem, we first relax ${M}$ to $\left( 0,+\infty \right)$. Considering that the area spectral efficiency is concave and increases with antenna number as shown in Remark 1, and that the energy consumption model grows linearly with antenna number, the optimal ${M}^*$ can be calculated by determining the first order derivative of the objective function. Then, we have the following result to obtain the optimal ${M}^*$.
		
		\textbf{Remark 3}: \textit{The optimal number of antenna ${M}^*=\max \left( {K}, \lceil \widetilde{{M} } \rceil \right) $, where $\widetilde{{M}}$ is the solution of the equation }
		\begin{equation} 
			\underset{\varOmega _{EE}\left( {M} \right)}{\underbrace{\frac{\partial \mathcal{R}}{\partial{M}}\left( \frac{P}{\theta}+{M} P_{cir}+{K} ^3P_{pre}+P_0 \right) -\mathcal{R} P_c}}=0 \label{EEP}
			\tag{49}
		\end{equation}
		\textit{There is a unique solution for ( (\ref{EEP}), which is located in $\left[ {K} -1,+\infty \right) $ and can be found by the bisection method.}

		\begin{proof}
			Similar to the proof of Theorem 4 in \cite{7442903}. The first order derivative of the left item of formula (\ref{EEP}) is calculated as follows:
			\begin{equation} 
				\frac{\partial \varOmega _{EE}\left( {M} \right)}{\partial {M}}=\frac{\partial ^2\mathcal{R}}{\partial {M} ^2}\left( \frac{P}{\theta}+{M} P_{cir}+{K} ^3P_{pre}+P_0 \right) 
				\tag{50}
			\end{equation}
			From the observation that $\mathcal{R}$ is a concave function of ${M}$ in Remark 1, we have $\frac{\partial ^2\mathcal{R}}{\partial {M} ^2}<0$. Thus, we can obtain $\frac{\partial \varOmega _{EE}\left( {M} \right)}{\partial {M}}<0$, and $\varOmega _{EE}$ is a decreasing function of ${M}$. Meanwhile, from the conclusion that $\mathcal{R}$ is an increasing function of ${M}$ and for ${K} -1$, we know that $\varOmega _{EE}\left( {K} -1 \right) >0$. While for ${M} \rightarrow \infty $, $\lim _{{M} \rightarrow \infty}\frac{\partial \mathcal{R}}{\partial {M}}=0$. Thus, $\lim _{{M} \rightarrow \infty}\varOmega _{EE}\left( {M} \right) <0$. 
			Since $\varOmega _{EE}\left( {M} \right)$ is a decreasing function of ${M}$, we conclude that there is a unique solution for (\ref{EEP}), which can be found in $\left[ {K} -1,+\infty \right)$ by using the bisection method. This concludes the proof, and we shall demonstrate it clearly in the simulation results section. 
		\end{proof}

		\section{Numerical Results} 
		In this section, we verify the validity of the described analytical results with simulations. Moreover, we demonstrate the impact of different system parameters on the sum-rate, area spectral efficiency and energy efficiency. In this sense, we consider a simulation environment in which the locations of BS and users are distributed as PPPs with varying densities. Additionally, the effects of power splitting ratio and number of antennas on the network performance are investigated in terms of sum-rate, area spectral efficiency, and energy efficiency. The numerical results are computed in the interference-limited environment in which $N=1$, $\sigma^2=0$ and $\alpha=4$.
		
		We have adopted the following methodology regarding Monte Carlo simulations for this study \cite{6516171, 7931666}. 
		\begin{itemize}
			\item[1)] Basically, a square of 1000 $m^2$ is considered to be the simulation area $A$ and the transmit power is $5W$.
			\item[2)] Based on the fact that BSs in the RSMA-enhanced multi-cell dense networks have a high density, we generate the locations of BSs using a Poisson distribution with density $\lambda_b=1/\left(\pi150^2\right)$. Also, noise is ignored.
			\item[3)] Within the simulation region, BSs are distributed uniformly. 
			\item[4)] Users are also generated in the same simulation area $A$ by Poisson distribution with density $\lambda_u=2\lambda_b$. 
			\item[5)] Each BS will serve $N$ groups of RSMA users.
			\item[6)] Independent channel gains are generated for all links between BSs and users.
			\item[7)] The $\mathrm{SIR}_{m}$ are computed as shown in (\ref{SINRC}) and (\ref{SINRP}). 
			\item[8)] Then the rate is computed as $\mathcal{R} _{m}=\mathrm{ln}\left( 1+\mathrm{SIR}_{m} \right)$
			\item[9)] In the end, the rate is calculated by repeating the steps above $m$ times and calculating $\mathcal{R}=\frac{1}{m}\sum_1^m{\mathcal{R} _m}$. In our simulations, we have considered $m=1000$.	
		\end{itemize}
		Fig.2 is our simulation results. As it is shown in this figure, the gap between the analytical results and simulation ones can be negligible.
		\begin{figure}
			\centering
			\includegraphics[width=3.5in]{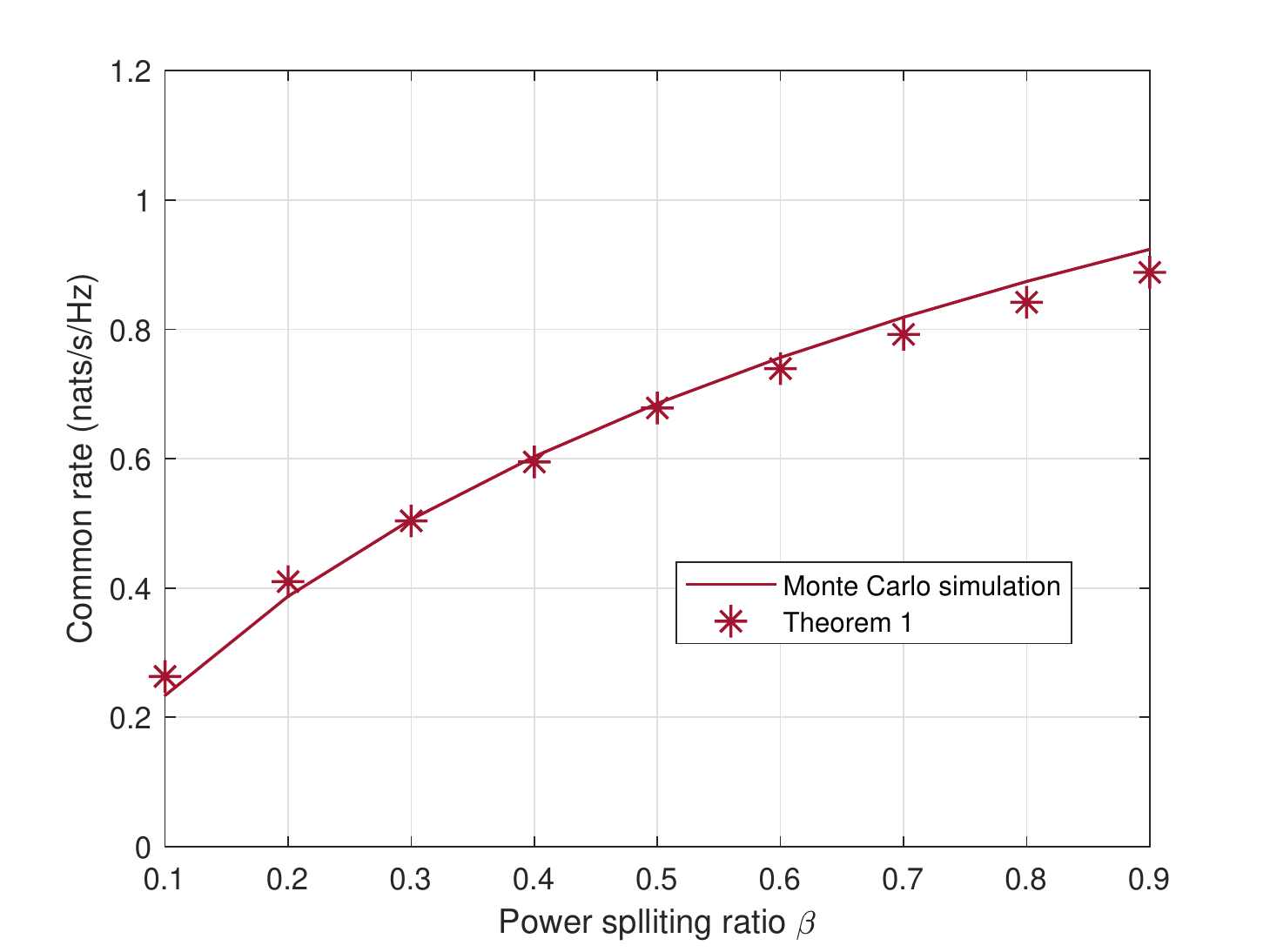}
			\caption{Sum-rate versus the power splitting ratio $\beta$ over Rayleigh fading in RSMA-enhanced multi-cell dense networks.}
		\end{figure}
		\subsection{Sum-Rate}
		Fig.3 demonstrates how different power splitting ratios affect the sum-rate of RSMA-enhanced multi-cell dense networks. It has shown that sum-rate is an increasing function of the power splitting ratio $\beta$, which is consistent with the conclusion in Remark 2. As a comparison, the sum-rates between RSMA and NOMA are evaluated under identical system conditions. Actually, multicasting performance is obtained by looking at $\beta=0$, and SDMA performance by looking at $\beta=1$. The sum-rate for RSMA, when compared to the curves with the same number of antennas, is higher than that of NOMA across a wide range of power splitting ratios $\beta$. Fig.3 illustrates the potential benefits of RSMA-enhanced multi-cell dense networks. It can be seen that there exists an optimal power splitting ratio which can maximize the sum-rate of users in the same RSMA group. For example, $\beta=0.1$ is the best choice for $M=4$ while $\beta=0.2$ might be better for $M=2$. 
		
		Fig.4 plots the common rate, sum-rate and sum of private rates versus the power splitting ratio and number of antennas. It is clear that common rate is a decreasing function of $\beta$ and the sum of private rates is an increasing function of $\beta$, which is consistent with the conclusion in Corollary 1. However, the growth of the sum of private rates is slower than the diminish of common rate. As a result, there exists an optimal $\beta$ which maximizes the sum-rate of RSMA-enhanced multi-cell dense networks. Moreover, we can see that the optimal power splitting ratio decreases with an increase in the number of antennas.

		\begin{figure}
			\centering
			\includegraphics[width=3.5in]{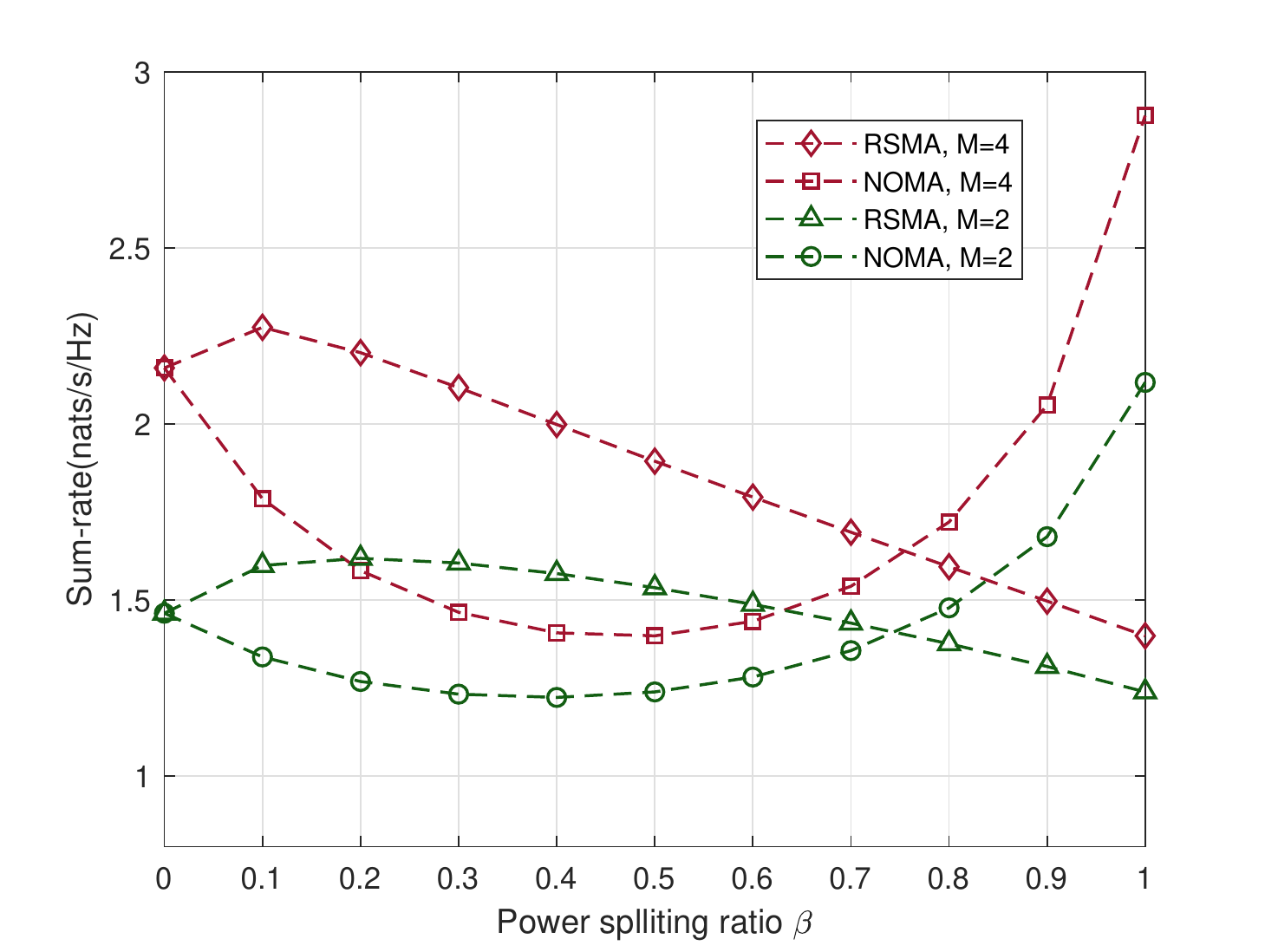}
			\caption{Sum-rate versus the power splitting ratio $\beta$ over Rayleigh fading in NOMA and RSMA-enhanced multi-cell dense networks for different antenna numbers.}
		\end{figure}
		
		Fig.5 shows the observation that follows from Remark 1, namely, the significant boost in the sum-rate with an increasing number of antennas for all MA techniques. As the result in \cite{7442903}, it is highly effective to deploy more antennas in multi-cell dense networks to improve the sum-rate for all MA technologies. However, the benefits of more antennas will diminish as the number of antennas increases. In dense networks, it is important to employ additional techniques besides multiple antennas so that interference can be minimized, and performance can be enhanced. Interested readers may refer to \cite{6415388} for a more detailed description of multiple antennas and how to optimize the deployment number. Nevertheless, the illustration consolidates that large number of antennas combined with RSMA has the greatest impact on enhancing performance over NOMA and SDMA in a wide range of power splitting ratio.
		
		A further illustration of the improvements in network performance is shown by Fig.6 and Fig.7, which shows the sum-rate gap between RSMA and NOMA calculated according to Proposition 2. It is obvious that RSMA has superior behavior to NOMA in a wide range of power splitting ratio. It can be also seen from Fig.6 and Fig.7 that the sum-rate gap between RSMA and NOMA increases with an increase in the number of antennas. On the other hand, in Fig.8, the sum-rate difference between RSMA and SDMA can also be found. From this result, we observe that the sum-rate for RSMA can outperform SDMA in a wide range of antenna numbers. Consequently, as shown in \cite{9451194}, RSMA can lead to a performance gains over NOMA and SDMA with lower receiver complexity. 
		\begin{figure}
			\centering
			\includegraphics[width=3.5in]{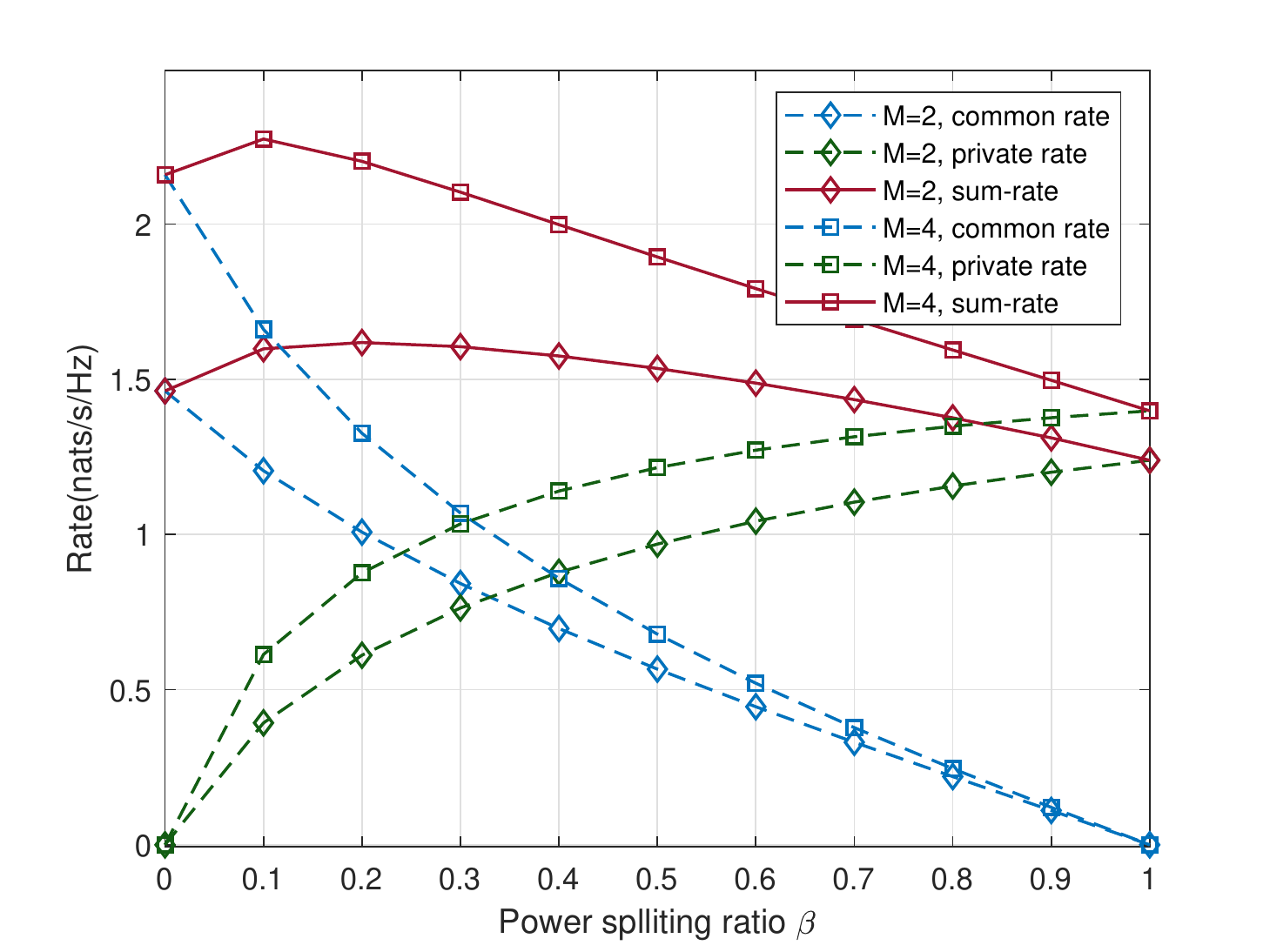}
			\caption{Sum-rate, common rate and sum of private rates versus the power splitting ratio $\beta$ over Rayleigh fading in RSMA-enhanced multi-cell dense networks.}
		\end{figure}
		
		\begin{figure}
			\centering
			\includegraphics[width=3.5in]{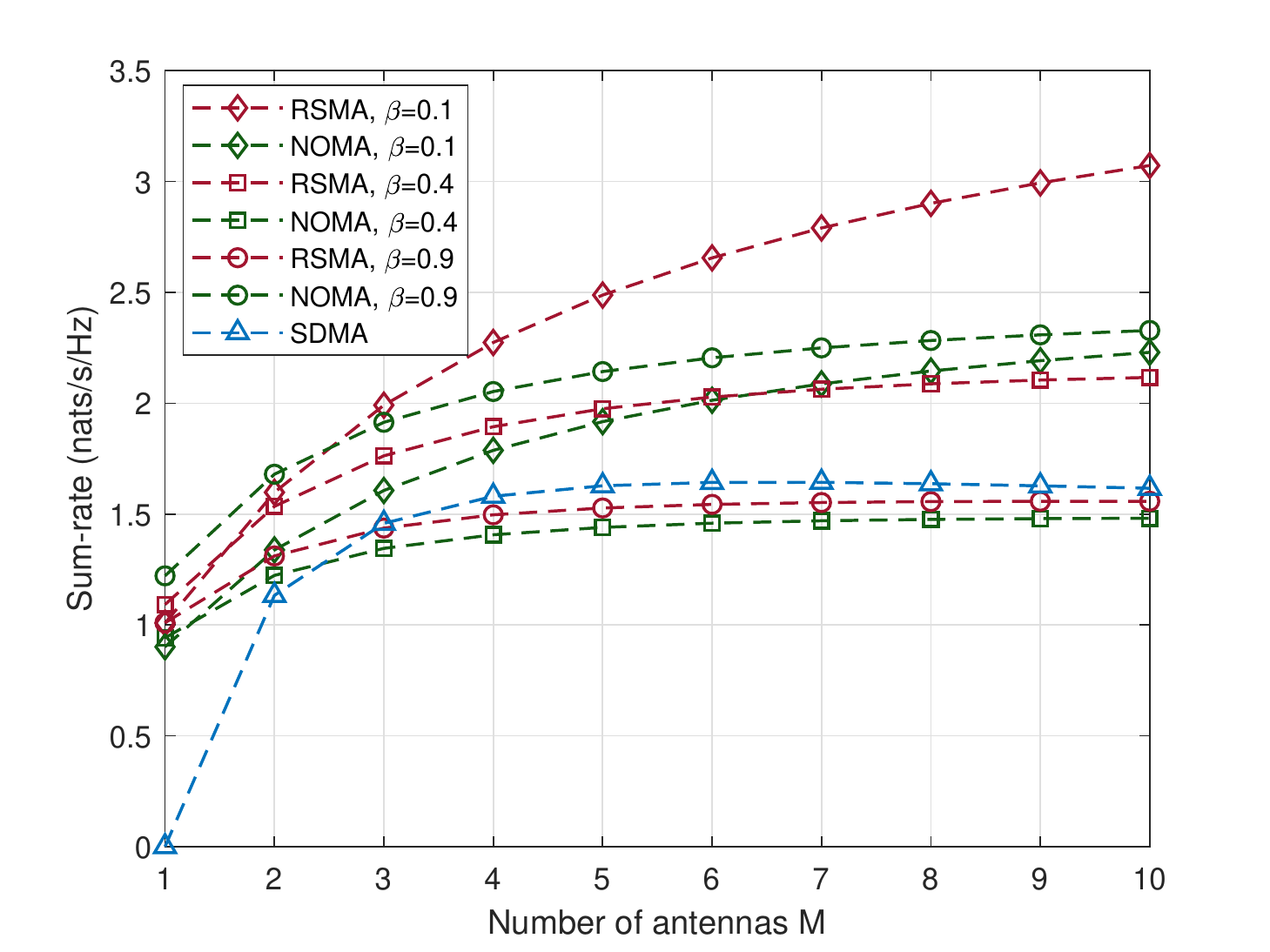}
			\caption{Sum-rate versus the number of antennas over Rayleigh fading in NOMA, SDMA and RSMA-enhanced multi-cell dense networks with different power splitting ratio $\beta$. }
		\end{figure}

		\begin{figure}
			\centering
			\includegraphics[width=3.5in]{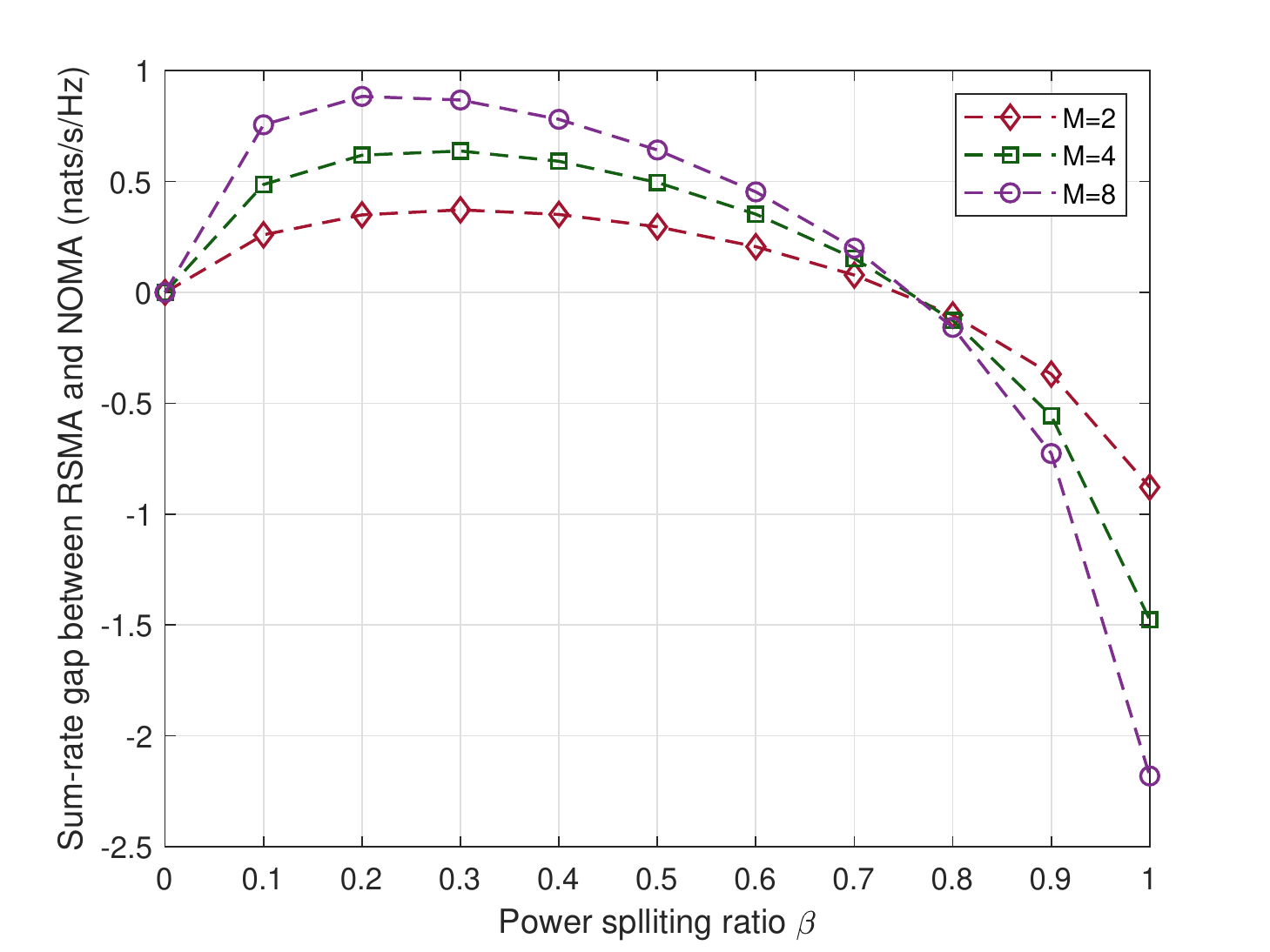}
			\caption{Sum-rate gap between RSMA and NOMA versus power splitting ratio $\beta$ in multi-cell dense networks for different antenna numbers $M$.}
		\end{figure}
		
		\begin{figure}
			\centering
			\includegraphics[width=3.5in]{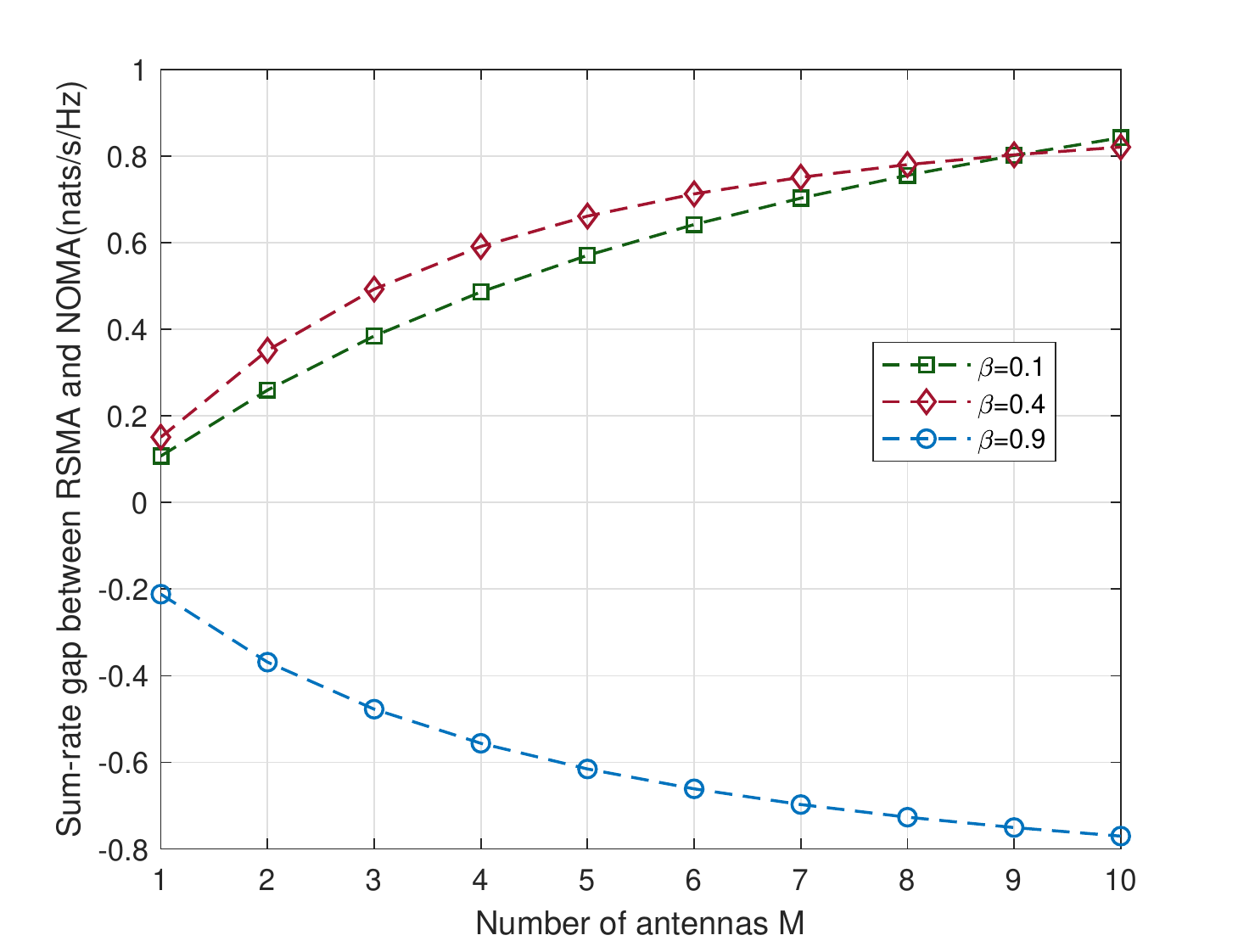}
			\caption{Sum-rate gap between RSMA and NOMA versus number of antennas $M$ in multi-cell dense networks for different power splitting ratio $\beta$.}
		\end{figure}
		
		\begin{figure}
			\centering
			\includegraphics[width=3.5in]{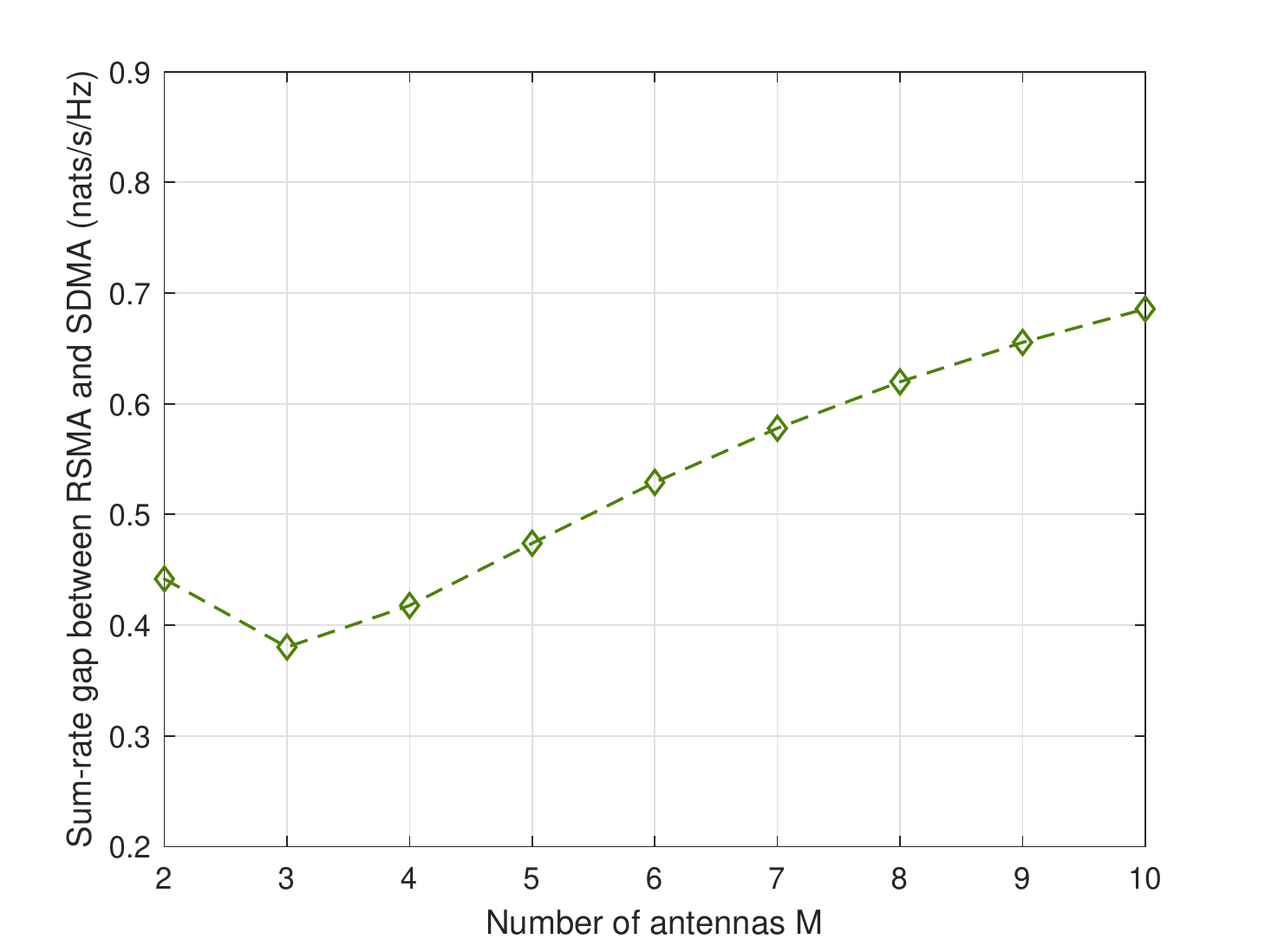}
			\caption{Sum-rate gap between RSMA and SDMA versus number of antennas $M$ in multi-cell dense networks.}
		\end{figure}
		
		\subsection{Area Spectral Efficiency and Energy Efficiency}
		Fig.9 plots the area spectral efficiency of multi-cell dense networks with RSMA, NOMA and SDMA versus the BS density. The analytical results are computed by (\ref{ASE}) in the interference-limited environment in which $\sigma^2=0$. As shown in the results, increasing BS density is an efficient way to boost the area spectral efficiency. Besides, the impacts of power splitting ratio and antenna number on the area spectral efficiency is consistent with those on sum-rate. This is because the area spectral efficiency is a linear function of the sum-rate.    
		
		Although expanding the number of antennas at the BS can enhance area spectral efficiency, the circuit power needed to operate each antenna contributes significantly to the energy consumption of the operation. To elucidate the trade-off between area spectral efficiency and energy consumption in terms of antenna number ${M}$, we consider the energy efficiency optimization problem which can be effectively calculated by (\ref{EE}). To the end, Fig.10 shows the effect of number of antennas on the energy efficiency. There exists an optimal value ${M}^*$ which maximizes the energy efficiency. This result gives us an insight in deploying appropriate antenna number from the perspective of energy efficiency. Furthermore, it also demonstrates that RSMA-enhanced multi-cell dense networks have a superior energy efficiency profile compared to NOMA and SDMA a wide variety of power splitting ratio. These observations emphasize the advantages of RSMA-enhanced multi-cell dense networks and provide insightful guidelines for design of practical dense networks.
		\begin{figure}
			\centering
			\includegraphics[width=3.5in]{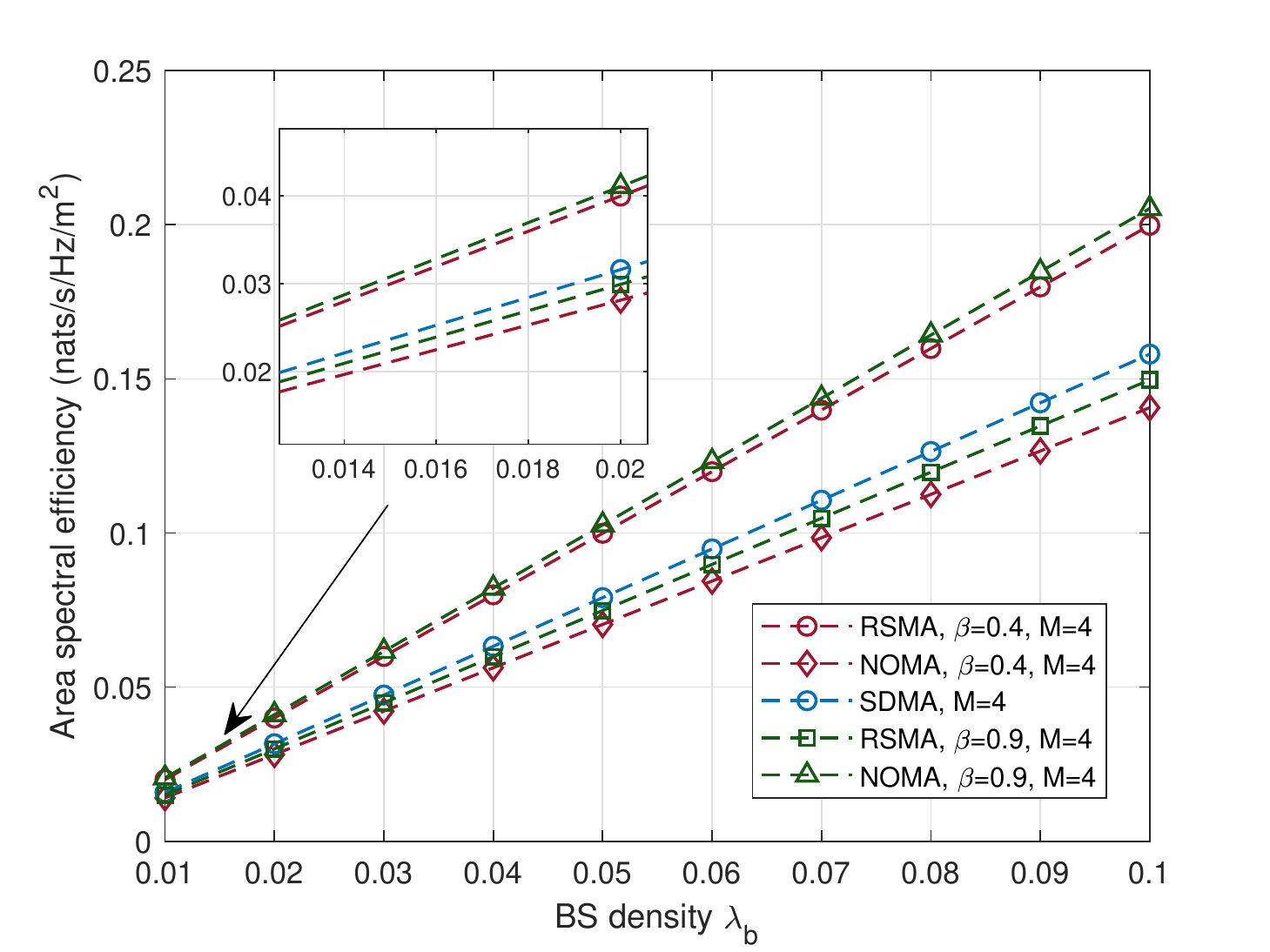}
			\caption{Area spectral efficiency versus the BS density $\lambda_b$ over Rayleigh fading in NOMA, SDMA and RSMA-enhanced multi-cell dense networks. }
		\end{figure}
		\begin{figure}
			\centering
			\includegraphics[width=3.5in]{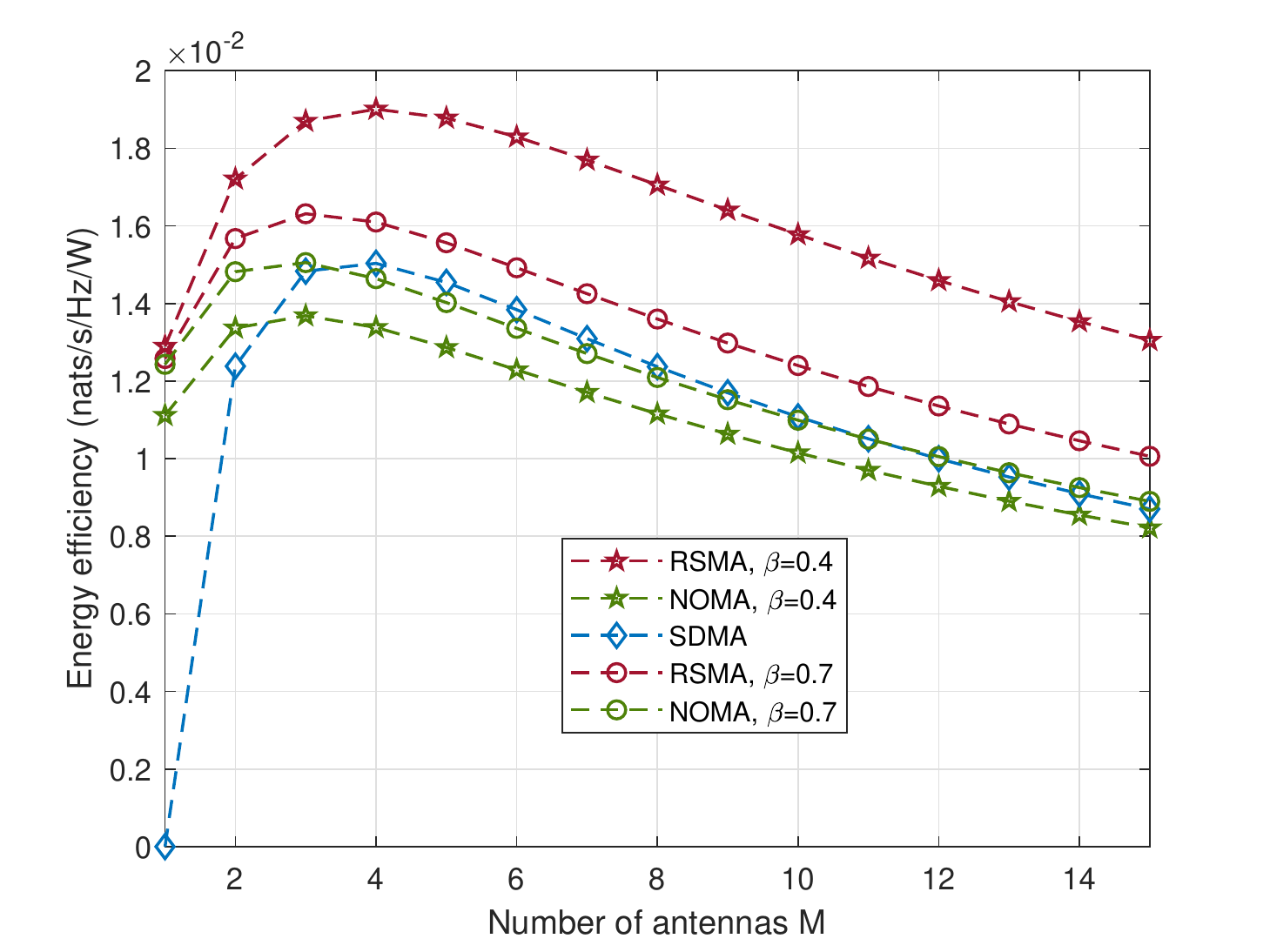}
			\caption{Energy efficiency versus the number of antennas over Rayleigh fading in NOMA, SDMA and RSMA-enhanced multi-cell dense networks. For the energy consumption model, the parameters are set as $P=5W, \theta =0.08, P_c=6.8W, P_{pre}=1.74, P_0=1.5$}
		\end{figure}

		\section{Conclusion}
		To conclude, an investigation of spatially average performance of the RSMA scheme in multi-cell dense networks has been presented in this paper. With stochastic geometry tools, we developed a mathematical framework for computing the performance metrics of the providing system by the MGF-based approach. Specifically, the analytical expressions for sum-rate, area spectral efficiency are derived based on this framework. Further, it was demonstrated that a different power splitting ratio between common stream and private streams will have a significant impact on the sum-rate and area spectral efficiency of the system. Moreover, it has shown that RSMA outperforms SDMA and NOMA in a wide range of power splitting ratio and deployment scenarios. In general, increasing the number of antennas at BSs improved network performance for all MA techniques, but RSMA technique could mostly result in greater benefits than NOMA and SDMA. At last, to pursue the energy optimal deployment strategy, the energy efficiency optimization problem was subsequently developed. For the future work, as multiple association in dense networks could be a good future focus for dealing with inter-cell interference, the combined application of this technique and RSMA could contribute to further improving the performance of multi-cell dense networks. It would also be useful to analyze the performance under imperfect CSIT, which typically happens when channel estimation is poor, and feedback is limited.

	\ifCLASSOPTIONcaptionsoff
	\newpage
	\fi
		
		\bibliographystyle{IEEEtran}
		\bibliography{IEEEexample}

	\end{document}